\documentclass[12pt, journal, draftclsnofoot, onecolumn]{IEEEtran}
\usepackage{graphicx}
\usepackage{amsmath}
\usepackage{amsfonts}
\usepackage{amssymb}
\usepackage{psfrag}
\usepackage{cite}
\usepackage{xcolor}
\usepackage{multirow}
\usepackage{subfigure}
\usepackage{url}
\newtheorem{theorem}{\underline{Theorem}}
\newtheorem{lemma}{\underline{Lemma}}
\newtheorem{remark}{\underline{Remark}}

\newtheorem{assumption}{\underline{Assumption}}
\newtheorem{problem}{\underline{Problem}}

\newtheorem{proposition}{\underline{Proposition}}
\newtheorem{definition}{\underline{Definition}}

\newcommand{\QED}{{\rm $\blacksquare$}}
\begin{document}
\title{\huge Design of Non-orthogonal and Noncoherent Massive MIMO for Scalable URLLC Beyond 5G}
\author{He Chen, Zheng Dong, Jian-Kang Zhang, and Branka Vucetic
\thanks{Part of the paper has been presented at 2019 IEEE International Conference on Industrial Cyber Physical Systems (ICPS) \cite{Dong2019icps}.

H. Chen is with the Department of Information Engineering, The Chinese University of Hong Kong, Shatin, NT, Hong Kong SAR, China (Email: he.chen@ie.cuhk.edu.hk).

Z. Dong is with the School of Information Science and Engineering, Shandong University, China (Email: zhengdong@sdu.edu.cn).

J.-K. Zhang is with the Department of Electrical and Computer Engineering, McMaster University, Hamilton, Ontario, L8S 4K1, Canada (Email:jkzhang@mail.ece.mcmaster.ca).

B. Vucetic are with the School of Electrical and Information Engineering, The University of Sydney, NSW 2006, Australia (Email: branka.vucetic@sydney.edu.au).

The first two authors contributed equally to this work.

}}

\maketitle
% ------------------------ abstract --------------------------------
\begin{abstract}
This paper is to develop and optimize a non-orthogonal and noncoherent massive multiple-input multiple-output (MIMO) framework towards enabling scalable ultra-reliable low-latency communications (sURLLC) in wireless systems beyond 5G. In this framework, the huge diversity gain associated with the large-scale antenna array in massive MIMO systems is leveraged to ensure ultrahigh reliability. To reduce the overhead and latency induced by the channel estimation process, we advocate the noncoherent communication technique which does not need the knowledge of instantaneous channel state information but only relies on the large-scale fading coefficients for information decoding. To boost the scalability of noncoherent massive MIMO systems, we enable the non-orthogonal channel access of multiple users by devising a new differential modulation scheme to assure that each transmitted signal matrix can be uniquely determined in the noise-free case and be reliably estimated in noisy cases when the antenna array size is scaled up. The key idea is to make the transmitted signals from multiple users be superimposed properly over the air such that when the sum-signal is correctly detected, the signal sent by each individual user can be uniquely determined. To further enhance the average error performance when the array antenna number is large, we propose a max-min Kullback-Leibler (KL) divergence-based design by jointly optimizing the transmitted powers of all users and the sub-constellation assignment among them. Simulation results show that the proposed design significantly outperforms the existing max-min Euclidean distance-based counterpart in terms of error performance. Moreover, our proposed approach also has a better error performance than the conventional coherent zero-forcing (ZF) receiver with orthogonal channel training, particularly for cell-edge users.
\end{abstract}

%----------------------------keywords--------------------------------------------------------------------

%--------------- section 1 ------------------------------------------------------------------------------
\section{Introduction}
%\subsection{Brief Introduction to IIoT}
The fifth generation (5G) of mobile networks is now rolling out globally and the newly deployed 5G infrastructure is expected to provide commercial services in 2020. Driven by the relentless growth of wireless data traffic over the past three decades, modern wireless communication systems (from 2G to nowadays 4G) have been consistently engineered and developed towards providing better mobile broadband services, with higher and higher data rates to subscribers. The trend is envisioned to continue in 5G as it will need to carry 10,000 times more traffic \cite{Nokiawhitepaper}. Besides, with the grand ambition of offering the connectivity to anything that may benefit from being connected, 5G cellular systems are envisaged to support two brand new service categories in addition to the conventional mobile broadband one: massive machine-type communication (mMTC) \cite{Dawy_WCM_2017} and ultra-reliable low-latency communication (URLLC) \cite{Soldani_NM_2018}.

The mMTC service refers to providing wireless connectivity for a massive number (tens of thousands) of low-cost and low-energy machine-type devices (MTDs) in a relatively large area. The mMTC can find potential applications in smart metering, smart agriculture, logistics, fleet management, etc. The traffic of these applications is characterized as massive yet sporadic small-packet transmissions which require the support of high spectrum efficiency and network scalability. Furthermore, the network maintenance cost can be huge due to the large amount of nodes. As such, ultra-high energy efficiency is demanded to achieve long battery lifetime of MTDs. On the other hand, URLLC is a service category not present in today's mobile systems, which targets for mission-critical applications requiring low end-to-end latency with high reliability. Examples are fault detection and isolation in power systems, detection and responses to hazardous road conditions, self-driven vehicles, remote surgery, smart factories, and augmented reality.

Nevertheless, with the continuing deployment of 5G cellular systems in practice, it gradually becomes clear that 5G is inadequate to fulfill the promised vision of being an enabler for the ``Internet of Everything", especially the most innovative URLLC part, due to its inherent limitations \cite{Saad2019}. While the enormous network capacity growth is achievable through conventional methods of moving to higher parts of the radio spectrum and network densifications, realizing URLLC will involve a departure from the underlying theoretical principles of wireless communications. More specifically, the coupling and contradictory requirements of low latency and high reliability render the design of URLLC systems a challenging task, since wireless channels are highly dynamic and are susceptible to fading, interference, blockage, and high pathloss, especially when there are lots of moving devices, metallic reflectors and electromagnetic radiation equipments~\cite{Durisi2016pieee,Popovski18,Ji2018WCM,Bennis2018pieee,Chen2018mag}. Such design challenges will be further escalated to provide the envisioned \emph{scalable URLLC} (sURLLC) services in wireless systems beyond 5G. As defined in \cite{Saad2019}, sURLLC will scale the 5G URLLC across the device dimension by seamlessly integrating 5G URLLC with legacy mMTC. In sURLLC, the augmented triple reliability-latency-scalability tradeoff will need to be carefully dealt with, which calls for a totally new design framework.

In wireless communications, channel diversity, which refers to a measure of transmitting multiple copies of the same  information through independent links along different time/frequency/spatial axises, is one of the most important techniques for boosting system reliability by effectively combating channel fading and interference~\cite{Tse05book}. As the diversity order increases, wireless channels will gradually become more stable and the chance of requesting retransmissions from the receiver side will correspondingly decrease~\cite{Popovski18}. However, in realizing sURLLC, time diversity is not preferred since achieving time diversity is at the cost of additional delay, especially in slow fading channels. Besides, delivering the information along distinct frequency channels will consume additional bandwidth, which is costly for the operations under 6\,GHz where the frequency band has already been over-crowded~\cite{ACMA15}. Thus, harnessing the spatial diversity by deploying multiple antennas at the transmitter and/or receiver side becomes the most appealing solution, and there have been extensive studies on various diversity technologies for conventional MIMO systems, see e.g.,~\cite{Tse03may, Marzetta2000tit} and references therein.

Recently, massive multiple-input multiple-output (MIMO) technology~\cite{Marzetta10}, which scales up the conventional MIMO by deploying a large number of antennas at the transmitter and/or receiver side, has been regarded as an indispensable {building} block for ensuring ultrahigh reliability~\cite{Popovski18}.
%The tradeoff  between latency, packet size, bandwidth, and finite-blocklength must be investigated.
%The massive MIMO techniques, where an excess number of transmitter and/or receiver antennas are exploited for channel diversity, has been a key enabling solution to realize the reliability requirement.
%In light of this paradigm shift, how to develop low-latency massive MIMO systems has become one of the most important research problems in the field.
Actually, massive MIMO has already been an integrating part of 5G communications for its great potential~\cite{Andrews14, Popovski14}. The main advantages of massive MIMO include high array gain, high spatial multiplexing gain and the immune to fast fading in rich scattering environments. The fluctuations of wireless channels can be averaged out in massive MIMO, and high reliability can be maintained for short packets without the need of strong channel coding. Despite the advantages mentioned above, the mechanisms of leveraging massive MIMO for realizing sURLLC are still largely unexplored.

%\subsection{Noncoherent Massive MIMO}
The reliability and latency gains associated with massive MIMO systems depend critically on the acquisition of the instantaneous channel state information (CSI)~\cite{Lu14Oct}.  In conventional communication systems, the estimation of instantaneous CSI is commonly achieved by transmitting {certain} known pilot symbols that are orthogonal to different users, where the channel estimation overhead is relatively low compared with the long data payload. However, the packets in sURLLC applications are typically very short and thus the overhead induced by channel estimation becomes non-negligible and will reduce the effective transmission rate significantly~\cite{Popovski16}. Moreover, sending pilots will cause significant delay in short-packet communications, especially over fast fading channels where the pilot symbols need to be dense in the time-frequency grid. As a matter of fact, obtaining instantaneous CSI is one of the most severe limiting factors to exploit the full potential of massive MIMO, where the latency introduced by the channel estimation in massive MIMO {constitutes} a major barrier for meeting the extreme delay requirement~\cite{Popovski18}. To reduce the latency in massive MIMO, the transmission protocol should depend on as less of the channel knowledge of small-scale fading as possible~\cite{Popovski18}.
%\emph{This section is dedicated to the exploitation of multiple antennas at the transmitter and receiver to support URLLC. At first, we need to establish the fact that the acquisition of the instantaneous CSI is one of the most severe limitations with respect to URLLC when exploiting multiple antennas; see Section VIII. This is because the CSI acquisition is a major protocol step in massive MIMO, impacting both the reliability and the latency. Taking this into account, we devise beamforming methods that rely mostly on the structure of the channel, that is, the direction of the propagation path. The information about small scale fading is exploited as little as possible. As the structure of the channel varies on a large scale basis, its acquisition is more robust to device mobility.}
Nevertheless, the knowledge of channel statistics remains crucial to provide high reliability requirements, especially when the precise knowledge of instantaneous CSI is not available. Noncoherent detection, where no instantaneous CSI is required, can be the key supporting asset for low-latency applications. It was showed in~\cite{Giuseppe2017arxiv} that noncoherent transmission is more energy efficient than pilot-assisted transmission schemes, even when the number of pilot symbols and their power are optimized.
%\emph{Acquisition of the CSI at the receiver is usually perceived as less critical than at the transmitter as the delay between channel estimation and data detection is short. However, extreme cases of mobility at the user side might require an alternative to coherent detection, especially if URLLC is the target. Hence, non-coherent detection methods can be an asset in that case. A particularly simple method [66] that greatly benefits from the presence of a massive number of antennas is based on energy detection at the uplink of a massive MIMO system. The principle is to send a single stream of data, collect and aggregate the energy from all antennas. Detection is performed based on the average channel energy across the antenna array, which tends to a deterministic  quantity for localized movements of the user and is therefore much more robust to user mobility than coherent detection. In addition, an efficient constellation design has been proposed in [67] that is able to benefit from the advantages of coherent communications at low mobility while switching to energy detection to ensure reliable communications at high mobility.}

We note that there have been considerable efforts on designing single-user noncoherent massive single-input multiple-output (SIMO) systems, see e.g.,~\cite{Goldsmith16twc,Popovski16tsp,Xie2019sys,Gao18iotj}, which demonstrated that simple energy-based modulation and detector can be sufficient for reliable detection by leveraging the massive number of antennas. By considering the single-user scenario, these work implicitly assumed that an orthogonal multiple access (OMA) mechanism (e.g., time-division multiple access) has been adopted at the data link layer to support the co-existence of multiple users. However, OMA mechanisms normally have poor scalability---the channel access latency scales up linearly as the number of end-devices increases, and thus are no longer suitable for the more challenging sURLLC applications with a large forecasted number of devices. One effective solution to address this scalability issue is to break the orthogonality of existing OMA protocols and empower a new non-orthogonal and noncoherent massive MIMO (nn-mMIMO) framework. It is worth mentioning here that non-orthogonal multiple access (NOMA) has recently received tremendous attentions from the mobile communication research community as a promising technology for 5G cellular system, see a recent comprehensive survey \cite{Dai2018CST}, where the primary goal of applying NOMA is to boost spectral efficiency and user fairness. Existing NOMA solutions along this research line normally require the estimation of instantaneous CSI such that the optimization of
 power allocation/control for different signal streams can be conducted at the transmitter side and successive interference cancellation can be implemented for detecting multiple users' signals at the receiver side. These NOMA solutions are thus not applicable anymore for nn-mMIMO enabled sURLLC applications.

Enabling NOMA in massive MIMO is in fact straightforward when the instantaneous CSI is available, which can be achieved by applying space domain multiple access (SDMA). However, how to empower the non-orthogonal access of multiple users at the same time in noncoherent massive MIMO systems becomes a non-trivial task as beamforming techniques can not be used anymore. Very recently, a new constellation domain-based NOMA methodology towards enabling nn-mMIMO has been developed in~\cite{Goldsmith16tit,Zhang2018JSAC,Xu2019ICC,chen2019icps}, which allows the simultaneous channel access of multiple devices at the data link layer without the availability of instantaneous CSI at the physical layer. However, all the designs in~\cite{Goldsmith16tit,Zhang2018JSAC,Xu2019ICC,chen2019icps} only considered one-shot communications (i.e., the received signal is decoded in a symbol-by-symbol manner). As such, the phase information of the transmitted signals is lost at the receiver side and thus only unipolar PAM constellations can be used, which largely limits the system reliability performance as the number of devices increases.

Towards enabling sURLLC, in this paper we develop a new nn-mMIMO framework that can perform joint noncoherent detection of the uplink signals from multiple devices over more than one time slots, where the transmitted signals are allowed to use the more robust QAM constellations. \emph{The main contributions of this paper are two-fold}:

Firstly, we apply a noncoherent maximum likelihood (ML) receiver which {relies} only on the second-order channel statistics and no instantaneous CSI is needed at either the transmitter or receiver sides. For the considered ML receiver, we systematically design a uniquely-factorable multiuser space-time modulation (UF-MUSTM) scheme to enable the concurrent transmission of multiple devices to a noncoherent receiver equipped with a large number of antennas. We further identify the necessary and sufficient conditions for the receiver to recover the transmitted signals from all users. Note that our design connects to the conventional space-time code design.
%A systematic design of UF-MUSTM based on the QAM constellation.
%Different from conventional noncoherent space-time code design, to enable the channel estimation, the condition that the codeword has a full row-rank can be guaranteed/verified beforehand by the transmitter. For example, such design can be realized by sending unitary codewords and using a general likelihood ratio test (GLRT) receiver~\cite{Varanasi01, Zhang11tit}. In our MUSTM, achieving full row-rank with dependent signals from different users is very challenging task and special design should be carried out to enable such property of the codewords.
Up to now, most of the existing space-time code designs, such as~\cite{Marzetta2000tit, Varanasi01, Aazhang03tit, Zhang11tit}, considered point-to-point MIMO systems, where all the transmitting antennas are connected to the same transmitter, and hence the transmitted information-carrying signals are accessible by all the antennas. However, in our considered UF-MUSTM based nn-mMIMO system, the signals transmitted from different users are not allowed to fully collaborate, which dramatically limits the codebook design. Particularly, the widely used unitary space-time code design is in general intractable for the considered multiuser massive MIMO system.

Secondly, we further optimize the proposed design framework by jointly design the constellations of multiple users. We note that the performance analysis for non-unitary codeword of MUSTM is extremely challenging, if not possible, as shown in~\cite{Varanasi01, Aazhang03tit}. Confronting such a challenge, we propose a max-min Kullback-Leibler (KL) {divergence}-based design criterion, based on which we jointly optimize the transmit powers of all users and the sub-constellation assignment among them. Note that the basic idea of this paper has been presented in the conference version \cite{Dong2019icps}, in which we only consider the simple scenario that all users adopt 4-QAM. In this paper, we have extended the design to the more general case that all users can adopt the larger QAM with not necessarily the same orders, which makes the optimization problem more complex and harder to resolve. We manage to resolve the formulated optimization problem in closed-form. Simulations are provided to demonstrate the superiority of the proposed design over the state-of-the-art benchmarking schemes.

{The remainder of this paper is organized as follows. In Sec. II, we describe the system model, the noncoherent detector, as well as the signal design. The design and optimization of the proposed UF-MUSTM framework is elaborated in Sec. III. Simulations are conducted and the corresponding results are discussed in Sec. IV. The conclusions are finally drawn in Sec. V. }
% {\bf{Notations}}:
%Matrices and column vectors are denoted by boldface characters with uppercase  (e.g., ${\mathbf A}$) and lowercase (e.g., ${\mathbf b}$), respectively.
%The transpose of ${\mathbf A}$ is denoted by ${\mathbf A}^T$ and the Hermitian transpose of ${\mathbf A}$ (i.e., the conjugate and transpose of ${\mathbf A}$) is denoted by ${\mathbf A}^H$.
%The $(m, n)$-th entry of ${\mathbf A}$ is denoted by $[{\mathbf A}]_{m,n}$.
%The columns of an $M\times N$ matrix ${\mathbf A}$ are denoted by ${\mathbf a}_1, {\mathbf a}_2, \cdots, {\mathbf a}_N$.
%The $m$-th entry of ${\mathbf b}$ is denoted by $b_m$ and $\left\|\mathbf b\right\|$ represents the Euclidean norm of $\mathbf b$. and the $n$-th diagonal entry of a matrix ${\mathbf A}$ is denoted by $[{\mathbf A}]_n=[{\mathbf A}]_{n,n}$.
%%Notation ${\mathbf A}^+$ stands for the pseudo-inverse of ${\mathbf A}$.
%We let $\otimes$ denote the Kronecker product and $j=\sqrt{-1}$.
%In addition, $\mathbf{a} \le \mathbf{b}$ means that $a_k \le b_k, k=1, 2, \ldots, n$ with $\mathbf{a} , \mathbf{b} \in \mathbb{R}^{n \times 1}$. We also let $\mathbf{0}$ and $\mathbf{1}$ be column vectors with all zero and all one entries of appropriate dimension, respectively, which will be clear from the context.

\section{System Model, Noncoherent Detector, and Signal Design}
\subsection{System Model and Noncoherent ML Detector}\label{sec:systemmodel}
We consider a massive {multiple-input multiple-output (MIMO)} system consisting of $K$ single-antenna users transmitting simultaneously to a base station (BS) with $M$ ($M\gg K$) receiving antennas on the same time-frequency grid. By using a discrete-time complex baseband-equivalent model, the received signal at the antenna array of BS in the $t$-th time slot\footnote{Each time slot refers to one symbol duration throughout this paper.}, defined as $\mathbf{y}_t=[y_{1,t},\ldots, y_{M,t}]^T$, can be expressed by
\begin{align*}
\mathbf{y}_t=\mathbf{H}\mathbf{x}_t +{\boldsymbol\xi}_t,
\end{align*}
where $\mathbf{x}_t=[ x_{1,t},\ldots, x_{K,t}]^T$ represents the transmitted signals from all $K$ users, ${\boldsymbol\xi}_t$ is an additive circularly-symmetric complex Gaussian (CSCG) noise vector with covariance $\sigma^2 \mathbf{I}_M$. We let $\mathbf{H}=\mathbf{G}\mathbf{D}^{1/2}$ denote the $M\times K$ complex channel matrix between the receiver antenna array and all users, where $\mathbf{G}$ characterizes the small-scale fading caused by local scattering while $\mathbf{D} ={\rm diag}\{\beta_1,\cdots, \beta_K\}$ with $\beta_k>0$ capturing the propagation loss due to distance and shadowing. All the entries of $\mathbf{G}$ are assumed to be i.i.d. complex Gaussian distributed with zero mean and unit variance. The channel coefficients are assumed to suffer from block fading which are quasi-static in the current block and change to other independent values in the next block with a channel coherence time $T_c \ge K$. We consider a space-time block modulation (STBM)~\cite{Aazhang03tit} scheme over $T$ time slots and the received signal vectors can be stacked together into a matrix form given by
\begin{align}\label{eqn:vecrecsignal}
\mathbf{Y}_T=\mathbf{H}\mathbf{X}_T +{\mathbf\Xi}_T,
\end{align}
where $\mathbf{Y}_T=[\mathbf{y}_1,\ldots, \mathbf{y}_T]$, $\mathbf{X}_T=[\mathbf{x}_1, \ldots, \mathbf{x}_T]$ and ${\mathbf \Xi}_T=[\boldsymbol{\xi}_1,\cdots, \boldsymbol{\xi}_T]$.

\begin{assumption}\label{assumption1}
 Throughout this paper, we adopt the following assumptions:
\begin{enumerate}
\item  The small scale channel fading matrix $\mathbf{G}$ is completely unknown to the BS and all the users, while the large scale fading matrix $\mathbf{D}$ is available at the BS that will be leveraged to optimize the system performance;
%as assumed in~\cite{Marzetta11}
\item The transmitted signals are subject to an instantaneous average power constraint\footnote{Note that our design can be directly extended to the case with peak power constraint.}: $\mathbb E \{|x_{k,t}|^2\} \le P_k$, $k=1,\ldots, K$, $t=1,\ldots, T$. For convenience, we assume that the users are labeled in an ascending order with $P_1\beta_1\le \ldots \le P_K \beta_K$. \hfill\QED
\end{enumerate}
\end{assumption}

In this work, we apply a noncoherent ML detector which is optimal for uniformly distributed discrete input signals in terms of error probability. We note that~\eqref{eqn:vecrecsignal} can be reformulated as $\mathbf{Y}_T^H=\mathbf{X}_T^H \mathbf{D}^{1/2}\mathbf{G}^H +{\boldsymbol \Xi}_T^H$. With the help of~\cite{Petersen12}, the vectorized form of the received signal can then be written as
\begin{align*}
\mathbf{y}={\rm vec}(\mathbf{Y}_T^H)=(\mathbf{I}_{M}\otimes \mathbf{X}_T^H \mathbf{D}^{1/2}){\rm vec}(\mathbf{G}^H) +{\rm vec}(\boldsymbol{\Xi}_T^H).
\end{align*}
As all the entries of $\mathbf{G}$ and $\boldsymbol \Xi$ are i.i.d. CSCG, we immediately have $\mathbb E[\mathbf{y}] =\mathbf{0}$, and the covariance matrix of $\mathbf{y}$ can be calculated as
$\mathbf{R}_{\mathbf{y}|\mathbf{X}_T}=\mathbb E[\mathbf{y}\mathbf{y}^H]
%&=\mathbb E\Big\{\big ((\mathbf{I}_M \otimes \mathbf{X}_T^H \mathbf{D}^{1/2}){\rm vec}(\mathbf{G}^H) +{\rm vec}(\boldsymbol{\Xi}^H)\big ) \big ( (\mathbf{I}_M \otimes \mathbf{X}_T^H \mathbf{D}^{1/2}){\rm vec}(\mathbf{G}^H) +{\rm vec}(\boldsymbol{\Xi}^H)\big )^H\Big\}\\
=\mathbf{I}_M \otimes (\mathbf{X}_T^H \mathbf{D}\mathbf{X}_T +\sigma^2 \mathbf{I}_T)$.
The conditional distribution of the received signal $\mathbf{y}$ at BS for any transmitted signal matrix $\mathbf{X}_T$ can then be given by $p({\mathbf{y}|\mathbf{X}_T})=\frac{1}{\pi^{K M}\det({\mathbf{R}_{\mathbf{y}|\mathbf{X}_T}})} \exp(-\mathbf{y}^H \mathbf{R}_{\mathbf{y}|\mathbf{X}_T}^{-1}\mathbf{y})$,
where $\mathbf{R}_{\mathbf{y}|\mathbf{X}_T}=\mathbf{I}\otimes (\mathbf{X}_T^H \mathbf{D}\mathbf{X}_T +\sigma^2 \mathbf{I}_T)$.
%Note that $\det(\mathbf{R}_{\mathbf{y}|\mathbf{X}_T})=\big(\det( \mathbf{X}_T^H \mathbf{D}\mathbf{X}_T +\sigma^2 \mathbf{I})\big)^M$ and  $\mathbf{y}^H \mathbf{R}_{\mathbf{y}|\mathbf{X}_T}^{-1}\mathbf{y}={\rm tr}\big\{\mathbf{Y}_T(\mathbf{X}_T^H\mathbf{D}\mathbf{X}_T +\sigma^2 \mathbf{I})^{-1} \mathbf{Y}_T^H \big\}$.
%Therefore, the received signal block $\mathbf{Y}_T$ conditioned on  transmitted information carrying matrix is given by:
%\begin{align*}
%f({\mathbf{Y}_T|\mathbf{X}_T})=\frac{1}{\pi^{KM}\det({\mathbf{R}_{\mathbf{y}|\mathbf{X}_T}})} \exp\Big\{-{\rm tr}\Big[\mathbf{Y}_T(\mathbf{X}_T^H \mathbf{D}\mathbf{X}_T +\sigma^2 \mathbf{I})^{-1} \mathbf{Y}_T^H \Big]\Big\},
%\end{align*}
The noncoherent ML detector can estimate the transmitted information carrying matrix from the received signal vector $\mathbf{y}$ by resolving the following optimization problem:
%\begin{align}
%\widehat{\mathbf{X}}_T={\arg\min}_{\mathbf{X}_T}~{\rm tr}\Big\{\mathbf{Y}_T(\mathbf{X}_T^H \mathbf{D}\mathbf{X}_T +\sigma^2 \mathbf{I})^{-1} \mathbf{Y}_T^H \Big\} +M \log\det( \mathbf{X}_T^H \mathbf{D}\mathbf{X}_T +\sigma^2 \mathbf{I}).
%\end{align}
\begin{align}\label{eqn:MLdetector}
\widehat{\mathbf{X}}_T={\arg\min}_{\mathbf{X}_T}~\mathbf{y}^H \mathbf{R}_{\mathbf{y}|\mathbf{X}_T}^{-1}\mathbf{y}+\log \det(\mathbf{R}_{\mathbf{y}|\mathbf{X}_T}).
\end{align}

From~\eqref{eqn:MLdetector}, we can observe that the detector relies on the sufficient statistics of the transmitted signal matrix $\mathbf{R}_{\mathbf{y}|\mathbf{X}_T}=\mathbf{I}\otimes (\mathbf{X}_T^H \mathbf{D}\mathbf{X}_T +\sigma^2 \mathbf{I}_T)$. The detailed discussion regarding the signal design is given in the following subsection.
% or more specifically,
%\begin{align}\label{eqn:correlationmatrix}
%\mathbf{X}_T^H \mathbf{D}\mathbf{X}_T=\begin{bmatrix}
%\mathbf{x}_1^H \mathbf{D} \mathbf{x}_1 & \mathbf{x}_1^H \mathbf{D} \mathbf{x}_2 &\cdots & \mathbf{x}_1^H \mathbf{D} \mathbf{x}_T \\
%\mathbf{x}_2^H \mathbf{D} \mathbf{x}_1 & \mathbf{x}_2^H \mathbf{D} \mathbf{x}_2 &\cdots &\mathbf{x}_2^H \mathbf{D} \mathbf{x}_T \\
%\vdots &\vdots &\ddots &\vdots\\
%\mathbf{x}_T^H \mathbf{D} \mathbf{x}_1 & \mathbf{x}_T^H \mathbf{D} \mathbf{x}_2 &\cdots &\mathbf{x}_T^H \mathbf{D} \mathbf{x}_T \\
%\end{bmatrix}.
%\end{align}
%From Remark~\ref{remark1}, as the initial attempt for solving this problem, we propose the following design consisting of $K$ time slots:
%%\begin{itemize}
%%\item We let $\mathbf x_1=[\sqrt{p_1}, \sqrt{p_2}, \ldots, \sqrt{p_K}]^T$ be known signals while $\mathbf x_2=[\sqrt{q_1} s_1, \sqrt{q_2} s_2, \ldots, \sqrt{q_K} s_K]^T$ for $t\ge 2$ are information bearing signals, where $s_k\in\mathcal{X}_{\rm BPSK, QPSK}$ for $k=1,\ldots, K$.
%%\item
%%\end{itemize}
%\begin{align}
%\mathbf{X}&=
%\begin{bmatrix}
%\sqrt{p_1} & \sqrt{q}_1 x_{1,2}  &\ldots & \sqrt{q}_1 x_{1,K}\\
%\sqrt{p_2} &  \sqrt{q}_2 x_{2,2} &\ldots & \sqrt{q}_2 x_{2,K}\\
%\vdots & \vdots &\ddots &\vdots\\
%\sqrt{p_K}  & \sqrt{q}_Kx_{K,2}  &  \ldots& \sqrt{q}_K x_{K,K}
%\end{bmatrix}.
%\end{align}
\subsection{Unique Identification of the Transmitted Signal Matrix}
%In this section, we reveal the necessary and sufficient condition for the reliable noncoherent multiuser massive SIMO communication system must have to recover $\mathbf{X}_T$.
In this subsection, we first identify what conditions the transmitted signal matrix must satisfy to ensure the unique identification of the transmitted signal matrix $\mathbf{X}_T$.
%and the channel coefficients $\mathbf{G}$.
We can observe from~\eqref{eqn:MLdetector} that, to achieve reliable communication between all users and the BS in the considered nn-mMIMO system, the BS must be able to uniquely determine each transmitted signal matrix ${\mathbf X}_T$ once $\mathbf{R}=\mathbf{X}_T^H \mathbf{D}\mathbf{X}_T$ has been identified, which can be formally stated as follows:
\begin{proposition}\label{proposition:UFCM}
	Any reliable communications for the multiuser nn-mMIMO system described in~\eqref{eqn:vecrecsignal} require that, for the transmitted signal matrix selected from ${\mathcal M}^{K \times T}\subseteq \mathbb C^{K \times T}$, if and only if there exist any two signal matrices $\mathbf{X}_T, \widetilde{\mathbf{X}}_T\in {\mathcal M}^{K \times T}$ satisfying $\mathbf{X}_T^H \mathbf{D}\mathbf{X}_T= \widetilde{\mathbf{X}}_T^H \mathbf{D}\widetilde{\mathbf{X}}_{T}$, then we have ${\mathbf{X}}_T=\widetilde{\mathbf{X}}_T$.\hfill\QED
\end{proposition}

The proof is provided in Appendix\ref{append:prop1}.

Inspired by Proposition~\ref{proposition:UFCM},  to facilitate our system design, we introduce the concept of uniquely-factorable multiuser space-time modulation (UF-MUSTM), the formal definition of which is given as follows:
\begin{definition}\label{def:udmustm}
A multiuser space-time modulation codebook $\mathcal{S}^{K\times T} \subseteq \mathbb C^{K \times T}$ is said to form a UF-MUSTM codebook if for any pair of codewords $\mathbf{S}, \widetilde{\mathbf{S}} \in \mathcal{S}^{K\times T}$ satisfying $\mathbf{S}^H \mathbf{S} = \widetilde{\mathbf{S}}^H \widetilde{\mathbf{S}}$, we have $\mathbf{S} =\widetilde{\mathbf{S}}$.	\hfill\QED
\end{definition}

Definition~\ref{def:udmustm} motivates us to design a UF-MUSTM codebook for the considered nn-mMIMO system. Therefore, our primary task in the rest of this paper is to develop a new framework for a systematic design of such UF-MUSTM ${\mathcal S}^{K\times T}$.

Before proceeding on, it is worth clarifying here that the UF-MUSTM code design is fundamentally different from existing noncoherent space-time code/modulation designs. Specifically,

\begin{itemize}
\item For the considered UF-MUSTM based nn-mMIMO system, the signals transmitted from different users cannot fully collaborate, and hence the widely used unitary space-time code design is intractable for the considered system. This is fundamentally different from most of conventional space-time code designs for point-to-point MIMO system, where unitary space-time code is feasible since all the transmitting antennas are connected to the same transmitter~\cite{Marzetta2000tit, Aazhang03tit}. Note that the error performance analysis of non-unitary codeword of MUSTM is very challenging as shown in~\cite{Varanasi01}.
\item Our design is asymptotically optimal when the number of BS antennas goes to infinity while keeping the transmitted power fixed. This is in contrast to most previous space-time coding designs which considered the asymptotic regime with the signal-to-noise ratio (SNR) going to infinity~\cite{Varanasi01, Zhang11tit,Aazhang03tit}.
\end{itemize}

\section{Design and Optimization of UF-MUSTM Framework}\label{SecIII}
%In this section, we propose a slot-by-slot detection receiver and then analyze the dominant term of its error performance when the antenna array size is large for general MUSTM scheme. Based on the result on the error performance analysis, we propose a UDCG based multiuser space-time modulation design where the codewords are guaranteed to have a full row-rank.
In this section, we present a UF-MUSTM framework with a slot-by-slot noncoherent ML detector. We find that when the number of receiving antennas increases, the pairwise error probability (PEP) between two codewords will be dominated by the Kullback-Leibler (KL) divergence between them. Motivated by this fact, a max-min KL divergence design criterion is proposed to optimize the transmit powers of all users and the sub-constellation assignment among them.
\subsection{KL Divergence between Transmitted Space-Time Modulation Codewords}
In practice, the computational complexity of the optimal noncoherent ML detector described in~\eqref{eqn:MLdetector} could be prohibitively high. Furthermore, the error performance analysis results available for the block transmission with general block size and ML receiver are too complicated to reveal insightful results for the input codeword design and the corresponding power allocation~\cite{Varanasi01}.
To resolve these problems as well as to reduce the receiver complexity, our main idea is to input a small block size into the ML receiver. If only one time slot is involved in the ML detector given in~\eqref{eqn:MLdetector}, i.e., when $T=1$, the correlation matrix $\mathbf{R}=\mathbf{X}_T^H \mathbf{D}\mathbf{X}_T$ degenerates into a real scalar $\mathbf{x}_1^H\mathbf{D}\mathbf{x}_1=\sum_{k=1}^K \beta_k |x_{k,1}|^2$, where the phase information of the transmitted symbols is lost and information bits from all users can only be modulated on the amplitudes of the transmitted symbols. Such a design typically has a low spectral efficiency~\cite{Goldsmith16twc, Goldsmith16tit, Popovski16tsp}. To improve the spectrum efficiency by allowing constellation with phase information being transmitted by all users, we need to feed the signals received in at least two time slots into the ML decoder~\cite{Varanasi01,Madhowtit,Aazhang03tit}.

As an initial attempt, in this paper we focus on a slot-by-slot ML detection over the first and $t$-th time slots, which is similar to the differential modulation with the hard-decision based noncoherent multiuser detection. More specifically, we let the transmitted signal matrix be  $\mathbf{X}_T=[\mathbf{x}_1, \ldots, \mathbf{x}_T]$. For detection purpose, we now stack the transmitted signal of the first and the $t$-th time slot as $\mathbf{X}_t=[\mathbf{x}_1, \mathbf{x}_t]$, and then make the decision on $\mathbf{Y}_t=[\mathbf{y}_1, \mathbf{y}_t]$ by using~\eqref{eqn:MLdetector}. For simplicity, we consider the transmitted signal from the first and second time slots, i.e., $\mathbf{X}_2=[\mathbf{x}_1, \mathbf{x}_2]$, hereafter, and the case of $\mathbf{X}_t$ follows similarly. We denote
 $\mathbf{R}_{\mathbf{y}|\mathbf{X}_2}=\mathbf{I}\otimes \mathbf{R}_2$, in which
\begin{align}\label{eqn:correlation2by2}
&\mathbf{R}_2=\mathbf{X}_2^H \mathbf{D}\mathbf{X}_2 +\sigma^2 \mathbf{I}_2
=\begin{bmatrix}
\mathbf{x}_1^H \mathbf{D} \mathbf{x}_1 +\sigma^2 & \mathbf{x}_1^H \mathbf{D} \mathbf{x}_2 \\
\mathbf{x}_2^H \mathbf{D} \mathbf{x}_1 & \mathbf{x}_2^H \mathbf{D} \mathbf{x}_2 +\sigma^2
\end{bmatrix}.
%=\begin{bmatrix}
%a & c\\
%c^* & b
%\end{bmatrix}.
\end{align}
%=\|\mathbf{x}_1\|_M^2+\sigma^2
%where $a=\mathbf{x}_1^H \mathbf{D} \mathbf{x}_1+\sigma^2$, $b=\mathbf{x}_2^H \mathbf{D} \mathbf{x}_2 +\sigma^2$, and $c=\mathbf{x}_1^H \mathbf{D} \mathbf{x}_2$, such that $ab>|c|^2$.
 By~\eqref{eqn:correlation2by2}, we have
\begin{align}
\mathbf{R}_2^{-1}
=\frac{1}{(\mathbf{x}_1^H \mathbf{D} \mathbf{x}_1+\sigma^2)(\mathbf{x}_2^H \mathbf{D} \mathbf{x}_2 +\sigma^2)-|\mathbf{x}_1^H \mathbf{D} \mathbf{x}_2|^2}
\begin{bmatrix}
\mathbf{x}_2^H \mathbf{D} \mathbf{x}_2 +\sigma^2 & -\mathbf{x}_1^H \mathbf{D} \mathbf{x}_2\\
-\mathbf{x}_2^H \mathbf{D} \mathbf{x}_1 & \mathbf{x}_1^H \mathbf{D} \mathbf{x}_1+\sigma^2
\end{bmatrix}.
\end{align}
As a consequence, the ML receiver can be reformulated as follows
%\begin{subequations}
\begin{align}\label{eqn:simplifiedMLreceiver}
\widehat{\mathbf{X}}_2&={\arg\min}_{\mathbf{X}_2}~\mathbf{y}^H \mathbf{R}_{\mathbf{y}|\mathbf{X}_2}^{-1}\mathbf{y}+\log \det(\mathbf{R}_{\mathbf{y}|\mathbf{X}_2})\nonumber\\
&={\arg\min}_{\mathbf{X}_2}~\frac{ (\mathbf{x}_1^H \mathbf{D} \mathbf{x}_1+\sigma^2) \|\mathbf{y}_2\|^2 + (\mathbf{x}_2^H \mathbf{D} \mathbf{x}_2 +\sigma^2) \|\mathbf{y}_1\|^2 -2 \Re(\mathbf{x}_1^H \mathbf{D} \mathbf{x}_2 \mathbf{y}_2^H \mathbf{y}_1)}{(\mathbf{x}_1^H \mathbf{D} \mathbf{x}_1+\sigma^2)(\mathbf{x}_2^H \mathbf{D} \mathbf{x}_2 +\sigma^2)-|\mathbf{x}_1^H \mathbf{D} \mathbf{x}_2|^2}\nonumber\\
&\qquad\qquad\qquad\qquad\qquad + M \ln \Big[(\mathbf{x}_1^H \mathbf{D} \mathbf{x}_1+\sigma^2)(\mathbf{x}_2^H \mathbf{D} \mathbf{x}_2 +\sigma^2)-|\mathbf{x}_1^H \mathbf{D} \mathbf{x}_2|^2\Big],
\end{align}
%\end{subequations}
where $\mathbf{y}_1$ and $\mathbf{y}_2$ are the received signal vectors in the first and second time slots, respectively.	
It can be observed that the diagonal entries in~\eqref{eqn:correlation2by2} are $\mathbf{x}_1^H\mathbf{D}\mathbf{x}_1=\sum_{k=1}^K \beta_k |x_{k,1}|^2$ and $\mathbf{x}_2^H\mathbf{D}\mathbf{x}_2=\sum_{k=1}^K \beta_k |x_{k,2}|^2$, in which the phase information is lost, while the off-diagonal term is $\mathbf{x}_1^H \mathbf{D} \mathbf{x}_2 =\sum_{k=1}^K \beta_k x_{k,1}^*x_{k,2}=\sum_{k=1}^K \beta_k |x_{k,1}||x_{k,2}|\exp\big (j\arg(x_{k,2}) -j\arg(x_{k,1})\big )$, indicating that we can transmit a known reference signal vector $\mathbf{x}_1$ in the first time slot and then transmit the information bearing signal vector $\mathbf{x}_2$ to imitate a ``differential-like" transmission~\cite{2001titZheng}.
%We now consider the pairwise error probability (PEP) of the noncoherent ML receiver, which can be formulated by
%\begin{align}
%&\Pr\big\{\mathbf X_2 \to \mathbf X_2'\big\} =\Pr\big\{\mathbf{y}^H \mathbf{R}_{\mathbf{y}|\widetilde{\mathbf X}_2}^{-1}\mathbf{y}+\log \det(\mathbf{R}_{\mathbf{y}|\widetilde{\mathbf X}_2}) < \mathbf{y}^H \mathbf{R}_{\mathbf{y}|\mathbf{X}_2}^{-1}\mathbf{y}+\log \det(\mathbf{R}_{\mathbf{y}|\mathbf{X}_2})\big\} \nonumber\\
%%&=\Pr\big\{ \mathbf{y}^H \big (\mathbf{R}_{\mathbf{y}|\widetilde{\mathbf X}_2}^{-1} -  \mathbf{R}_{\mathbf{y}|\mathbf{X}_2}^{-1}\big )\mathbf{y} <  \log\det  (\mathbf{R}_{\mathbf{y}|\mathbf{X}_2} \mathbf{R}_{\mathbf{y}|\widetilde{\mathbf X}_2}^{-1}) \big\}\\
%&=\Pr\big\{ \mathbf{y}^H \big (\mathbf{R}_{\mathbf{y}|\widetilde{\mathbf X}_2}^{-1} -  \mathbf{R}_{\mathbf{y}|\mathbf{X}_2}^{-1}\big )\mathbf{y} <  M\log\det  (\mathbf{X}_2^H \mathbf{D}\mathbf{X}_2 +\sigma^2 \mathbf{I}_2) (\widetilde{\mathbf X}_2^H \mathbf{D}\widetilde{\mathbf X}_2 +\sigma^2 \mathbf{I}_2)^{-1} \big\}.
%\end{align}
The exact PEP is extremely hard to evaluate for the matrix $\mathbf{X}_2$ given above. %~\footnote{\comment{Please refer to the supplemental material for the detailed discussion.}}.
Moreover, the exact expression for the PEP does not seem to be tractable for further optimization. Inspired by the Chernoff-Stein Lemma, when the number of receiver antennas $M$ goes to infinity, the PEP will goes to zero exponentially where the exponent is determined by the KL divergence~\cite{Aazhang03tit}. Hence, in this paper we propose to use the KL divergence between the conditional distributions of the received signals for different inputs as the design criterion thanks to its mathematical tractability.

We now derive the KL divergence between the received signals induced by the  transmitted signals matrices $\mathbf{X}_2=[\mathbf{x}_1, \mathbf{x}_2]$ and $\widetilde{\mathbf{X}}_2=[\tilde{\mathbf{x}}_1, \tilde{\mathbf{x}}_2]$, which is also the expectation of the likelihood function between two received signal vectors. Essentially, the likelihood function between the received signal vectors corresponding to the two transmitted signals converge in probability to the KL-divergence as the number of receiver antennas increases~\cite{Aazhang03tit}. More specifically, the KL-divergence between the received signals corresponding to the transmitted matrix $\mathbf{X}_2$ and $\widetilde{\mathbf{X}}_2$ can be calculated as
\begin{subequations}
	\begin{align*}
	&\mathcal{D}_{\rm KL}^{(M)}(\mathbf{X}_2 ||\widetilde{\mathbf{X}}_2)=\mathbb E_{f({\mathbf{y}|\mathbf{X}_2})} \bigg\{ \ln \Big ( \frac{f({\mathbf{y}|\widetilde{\mathbf{X}}_2})}{f({\mathbf{y}|\mathbf{X}_2})} \Big ) \bigg\}\\
	&=\mathbb E_{f({\mathbf{y}|\mathbf{X}_2})} \bigg\{ \ln \Big (\frac{\det({\mathbf{R}_{\mathbf{y}|\mathbf{X}_2}}) }{\det({\mathbf{R}_{\mathbf{y}|\widetilde{\mathbf{X}}_2}})}\Big ) +\Big ( \mathbf{y}^H \mathbf{R}_{\mathbf{y}|\widetilde{\mathbf{X}}_2}^{-1}\mathbf{y} -\mathbf{y}^H \mathbf{R}_{\mathbf{y}|\mathbf{X}_2}^{-1}\mathbf{y}\Big )\bigg\}\\
	&=\mathbb E_{f({\mathbf{y}|\mathbf{X}_2})} \bigg\{ {\rm tr}\Big (\big ( \mathbf{R}_{\mathbf{y}|\widetilde{\mathbf{X}}_2}^{-1} - \mathbf{R}_{\mathbf{y}|\mathbf{X}_2}^{-1}\big )\mathbf{y}\mathbf{y}^H \Big )\bigg\}+\ln \Big ( \frac{\det({\mathbf{R}_{\mathbf{y}|\mathbf{X}_2}}) }{\det({\mathbf{R}_{\mathbf{y}|\widetilde{\mathbf{X}}_2}})}\Big )\\
	&= {\rm tr}\Big (\big (\mathbf{R}_{\mathbf{y}|\widetilde{\mathbf{X}}_2}^{-1} - \mathbf{R}_{\mathbf{y}|\mathbf{X}_2}^{-1}\big )\mathbf{R}_{\mathbf{y}|\mathbf{X}_2} \Big )+\ln \Big (\frac{\det({\mathbf{R}_{\mathbf{y}|\mathbf{X}_2}}) }{\det({\mathbf{R}_{\mathbf{y}|\widetilde{\mathbf{X}}_2}})}\Big )\\
	%&=M {\rm tr} \big\{ (\mathbf{X}_2^H \mathbf{D}\mathbf{X}_2+\sigma^2 \mathbf{I}_2)(\widetilde{\mathbf{X}}_2^H \mathbf{D}\widetilde{\mathbf{X}}_2 +\sigma^2 \mathbf{I}_2)^{-1} \big\} \\
	%&\qquad -M \ln\Big\{ \det\big ((\mathbf{X}_2^H \mathbf{D}\mathbf{X}_2+\sigma^2 \mathbf{I}_2)(\widetilde{\mathbf{X}}_2^H \mathbf{D}\widetilde{\mathbf{X}}_2 +\sigma^2 \mathbf{I}_2)^{-1}\big )\Big\} -2M\\
	&=M\, \mathcal{D}_{\rm KL}(\mathbf{X}_2 ||\widetilde{\mathbf{X}}_2),
	\end{align*}
\end{subequations}
in which
\begin{align}
\mathcal{D}_{\rm KL}(\mathbf{X}_2 ||\widetilde{\mathbf{X}}_2)&= {\rm tr} \big\{ (\mathbf{X}_2^H \mathbf{D}\mathbf{X}_2+\sigma^2 \mathbf{I}_2)(\widetilde{\mathbf{X}}_2^H \mathbf{D}\widetilde{\mathbf{X}}_2 +\sigma^2 \mathbf{I}_2)^{-1} \big\} \nonumber\\
&\qquad - \ln\Big\{ \det\big ((\mathbf{X}_2^H \mathbf{D}\mathbf{X}_2+\sigma^2 \mathbf{I}_2)(\widetilde{\mathbf{X}}_2^H \mathbf{D}\widetilde{\mathbf{X}}_2 +\sigma^2 \mathbf{I}_2)^{-1}\big )\Big\} -2.\label{eqn:KLdistanece}
\end{align}
We can observe from the above expression that $\mathcal{D}_{\rm KL}(\mathbf{X}_2 ||\widetilde{\mathbf{X}}_2)$ is actually the KL-divergence when there is only one receiving antenna. Due to the assumption of the independence of channel coefficients, and the KL divergence with $M$ antennas $\mathcal{D}_{\rm KL}^{(M)}(\mathbf{X}_2 ||\widetilde{\mathbf{X}}_2)$ is $M$ times of $\mathcal{D}_{\rm KL}(\mathbf{X}_2 ||\widetilde{\mathbf{X}}_2)$.
%i.e., $\mathcal{D}_{\rm KL}(\mathbf{X}_2 ||\mathbf{X}_t')=M \mathcal{D}_{\rm KL}(\mathbf{X}_2 ||\widetilde{\mathbf X}_2)$.

\subsection{QAM Division Based Multiuser Space-Time Modulation}
%\subsection{The Uniquely Decomposable Constellation Group}
The main objective of this subsection is to develop a new QAM division based MUSTM design framework for the considered {nn-mMIMO} system. The design is built upon the uniquely decomposable constellation group (UDCG) originally proposed in~\cite{dong16jstsp,Dong2017isit} for the commonly used spectrally efficient QAM signaling.
%To make this paper self-sufficient,
We now introduce the definition of UDCG as follows:
\begin{definition}\label{def:audcg}
	A group of constellations $\{\mathcal{X}_k\}_{k=1}^K$ form a UDCG, denoted by $\big\{ \sum_{k=1}^K x_k: x_k \in \mathcal{X}_k\big\}=\uplus_{k=1}^K \mathcal{X}_k = \mathcal{X}_1 \uplus \ldots \uplus \mathcal{X}_K$, if there exist two groups of $x_k, \tilde x_k \in \mathcal{X}_k$ for $k=1, \cdots, K$ such that $\sum_{k=1}^K x_k =\sum_{k=1}^K \tilde x_k$, then we have $x_k =\tilde x_k$ for $k=1, \cdots, K$.~\hfill\QED
\end{definition}

As PAM and QAM constellations are commonly used in modern digital communications, which have simple geometric structures, we now give the following construction of UDCG.
\begin{lemma}\label{lemma:UDCG} The UDCG with PAM and QAM constellations can be constructed as follows:

1) \emph{\underline{{UDCG} with PAM Constellation}}:
	For two given positive integers $K$ and $N$ ($N \ge K$) and nonnegative integer sequence $\{N_k\}_{k=1}^K$ satisfying $\sum_{k=1}^K N_k=N$, a $2^N$-ary PAM constellation $\mathcal{G} = \{\pm (m-\frac{1}{2})d : m =1,\ldots, 2^{N -1}\}$, with $d$ being the minimum Euclidean distance between the constellation points, can be uniquely decomposed into the sum of $K$ sub-constellations $\{{\mathcal X}_k\}_{k=1}^K$ denoted by $\mathcal{G} = \uplus_{k=1}^K \mathcal{X}_k$, where
	$\mathcal{X}_1 = \big\{\pm (m-\frac{1}{2})d\big\}_{m=1}^{2^{N_1 -1}}$,
	and
	$\mathcal{X}_k =
	\big\{\pm (m-\frac{1}{2}) \times 2^{\sum_{\ell=1}^{k-1} N_\ell }d\big\}_{m=1}^{2^{N_k -1}}$  for $k \ge 2$.
	
2) \emph{\underline{{UDCG} with QAM Constellation}}:
	For two positive integers $K$ and $N =N_I+N_Q$ ($N\ge K$), with $N_I$ and $N_Q$ being nonnegative integers that denote the sizes of the in-phase and quadrature components, respectively. Let $\{N_{I,k}\}_{k=1}^K$ and $\{N_{Q,k}\}_{k=1}^K$ denote two given nonnegative integer sequences satisfying $N_I =\sum_{k=1}^K N_{I,k}$ and $N_Q =\sum_{k=1}^K N_{Q,k}$ with $N_k=N_{I,k}+N_{Q,k}>0$. Then, there exists a PAM and QAM mixed constellation $\mathcal{Q} = \uplus_{k =1}^K \mathcal{X}_k$ such that  $\mathcal{X}_k= \mathcal{X}_{I,k} \uplus j\mathcal{X}_{Q,k}$, with $j\mathcal{X}_{Q,k} =\{jx:x\in\mathcal{X}_{Q,k}\}$, where $\mathcal{Q}_I =\uplus_{k=1}^K \mathcal{X}_{I,k}$ and $\mathcal{Q}_{Q} =\uplus_{k=1}^K \mathcal{X}_{Q,k}$ are two PAM UDCGs according to the rate allocation $\{N_{I,k}\}_{k=1}^K$ and $\{N_{Q,k}\}_{k=1}^K$, respectively.	~\hfill\QED
\end{lemma}

With the concept of UDCG, we are now ready to propose a QAM division based UF-MUSTM for the considered nn-mMIMO system with a noncoherent ML receiver given in~\eqref{eqn:simplifiedMLreceiver}. The structure of each transmitted signal matrix is given by $\mathbf{X}_{2}=[\mathbf{x}_1, \mathbf{x}_2]=\mathbf{D}^{-1/2}\mathbf{\Pi} {\mathbf S}_2$, in which
\begin{align}\label{eqn:sigmtx}
\mathbf{S}_2&= [\mathbf{s}_1, \mathbf{s}_2]= \begin{bmatrix}
	\frac{1}{\sqrt{p_1}}& \sqrt{p_1} s_{1}\\
	\frac{1}{\sqrt{p_2}}& \sqrt{p_2} s_{2}\\
	\vdots & \vdots\\
	\frac{1}{\sqrt{p_K}} & \sqrt{p_K} s_{K}
	\end{bmatrix}.
\end{align}

In our design, the diagonal matrix $\mathbf{D}^{-1/2}$ is used to compensate for the different large scale fading among various users. The vector $\mathbf{p}=[p_1, \ldots, p_K]$ is introduced to adjust the relative transmitting powers between all users, and $\mathbf{s}=[s_1, \ldots, s_K]$ is the information-carrying vector. The instantaneous power constraint can be given by $\mathbb E\{|x_{k,t}|^2\} \le P_k$, $k=1,\ldots, K$ and $t=1,2$.
We let $s_k\in \mathcal{X}_k$, where all $\mathcal{X}_k$'s constitute a UDCG with sum-QAM constellation $\mathcal{Q}$ such that $\mathcal{Q} =\uplus_{k=1}^K \mathcal{X}_k$ as defined in Lemma~\ref{lemma:UDCG}. The rate allocation between the $K$ users are based on the sum-decomposition such that $\sum_{k=1}^K N_k=N$ in which $N_k=N_{I,k}+N_{Q,k} =\log_2 (|\mathcal{X}_k|)$ denotes the bit rate of the user constellation $\mathcal{X}_k$.
%In other words, the transmitted signal from all the users in the $t$-th ($t\ge 2$) time slot  belongs to a UDCG.
The matrix $\mathbf{\Pi}=[\mathbf{e}_{\pi{(1)}}, \ldots, \mathbf{e}_{\pi(K)}]^T$ is a permutation matrix, where  $\mathbf{e}_k$ denotes a standard basis column vector of length $K$ with 1 in the $k$-th position and 0 in other positions. $\pi:\{1,\ldots,K\} \to \{1,\ldots, K\}$ is a permutation over $K$ elements characterized by
$\begin{pmatrix}1 &2 &\ldots & K\\\pi(1) &\pi(2)&\ldots&\pi(K)\end{pmatrix}$. We also let  $\pi^{-1}:\{1,\ldots,K\} \to \{1,\ldots, K\}$ be a permutation such that $\pi^{-1}(\pi(k))=k$ for $k=1,\ldots, K$. From the above definition, we immediately have $\mathbf{\Pi}^T\mathbf{\Pi}=\mathbf{I}_K$.

For the transmitted signal matrix $\mathbf{X}_{2}$, we have the following desired properties:
\begin{proposition}\label{prop:udcg}
Consider $\mathbf{X}_{2}=\mathbf{D}^{-1/2}{\mathbf \Pi}{\mathbf S}_2$ and $\widetilde{\mathbf{X}}_{2}=\mathbf{D}^{-1/2} \mathbf{\Pi}\widetilde{\mathbf S}_2$, where ${\mathbf S}_2$ and $\widetilde{\mathbf S}_2$ belong to ${\mathcal S}^{K\times 2}$  as described in Definition~\ref{def:udmustm}. If ${\mathbf X}^H_2{\mathbf D}{\mathbf X}_2=\widetilde{\mathbf X}^H_2{\mathbf D}\widetilde{\mathbf X}_2$, then we have ${\mathbf X}_2=\widetilde{\mathbf X}_2$.
	\hfill\QED
\end{proposition}

The proof of Proposition~\ref{prop:udcg} is given in~Appendix\ref{append:prop2}.

\subsection{User-constellation Assignment and Power Allocation}

To further enhance the system reliability performance, we now optimize the user-constellation assignment $\pi$ and power allocation vector $\mathbf{p}$ for the proposed nn-mMIMO framework. For the transmitted signal matrix considered in~\eqref{eqn:sigmtx}, we have
%\begin{align*}
%&\mathbf{X}_2^H \mathbf{D}\mathbf{X}_2 +\sigma^2 \mathbf{I}_2
%%=\begin{bmatrix}
%%\mathbf{x}_1^H \mathbf{D} \mathbf{x}_1 +\sigma^2 & \mathbf{x}_1^H \mathbf{D} \mathbf{x}_2 \\
%%\mathbf{x}_2^H \mathbf{D} \mathbf{x}_1 & \mathbf{x}_2^H \mathbf{D} \mathbf{x}_2 +\sigma^2
%%\end{bmatrix}
%=\begin{bmatrix}
%  \mathbf{s}_1^H \mathbf{s}_1 +\sigma^2 & \mathbf{s}_1^H \mathbf{s}_2 \\
%\mathbf{s}_2^H \mathbf{s}_1 & \mathbf{s}_2^H \mathbf{s}_2 +\sigma^2
%\end{bmatrix}
%=\begin{bmatrix}
%a & c\\
%c^* & b
%\end{bmatrix},\\
%&\widetilde{\mathbf{X}}_2^H \mathbf{D}\widetilde{\mathbf{X}}_2 +\sigma^2 \mathbf{I}_2
%=\begin{bmatrix}
%\mathbf{s}_1^H \mathbf{s}_1 +\sigma^2 & \mathbf{s}_1^H \tilde{\mathbf{s}}_2 \\
%\tilde{\mathbf{s}}_2^H \mathbf{s}_1 & \tilde{\mathbf{s}}_2^H \tilde{\mathbf{s}}_2 +\sigma^2
%\end{bmatrix}=\begin{bmatrix}
%a & \tilde{c}\\
%\tilde{c}^* & \tilde{b}
%\end{bmatrix},
%\end{align*}
\begin{align}\label{eqn:correlationmatrix}
&\mathbf{X}_2^H \mathbf{D}\mathbf{X}_2 +\sigma^2 \mathbf{I}_2
=\begin{bmatrix}
\mathbf{s}_1^H \mathbf{s}_1 +\sigma^2 & \mathbf{s}_1^H \mathbf{s}_2 \\
\mathbf{s}_2^H \mathbf{s}_1 & \mathbf{s}_2^H \mathbf{s}_2 +\sigma^2
\end{bmatrix}
=\begin{bmatrix}
\sum_{k=1}^K {1}/{p_k}+\sigma^2 & \sum_{k=1}^K s_k\\
\sum_{k=1}^K s_k^* & \sum_{k=1}^K p_k |s_k|^2 +\sigma^2
\end{bmatrix},\nonumber\\
&\widetilde{\mathbf{X}}_2^H \mathbf{D}\widetilde{\mathbf{X}}_2 +\sigma^2 \mathbf{I}_2
=\begin{bmatrix}
\mathbf{s}_1^H \mathbf{s}_1 +\sigma^2 & \mathbf{s}_1^H \tilde{\mathbf{s}}_2 \\
\tilde{\mathbf{s}}_2^H \mathbf{s}_1 & \tilde{\mathbf{s}}_2^H \tilde{\mathbf{s}}_2 +\sigma^2
\end{bmatrix}=\begin{bmatrix}
\sum_{k=1}^K {1}/{p_k}+\sigma^2 & \sum_{k=1}^K \tilde{s}_k\\
\sum_{k=1}^K \tilde{s}_k^* & \sum_{k=1}^K p_k |\tilde{s}_k|^2 +\sigma^2
\end{bmatrix}.
\end{align}
%in which $a=\sum_{k=1}^K \frac{1}{p_k}+\sigma^2$, $b=\sum_{k=1}^K p_k |s_k|^2 +\sigma^2$, $\tilde{b}=\sum_{k=1}^K p_k |\tilde{s}_k|^2 +\sigma^2$, $c=\sum_{k=1}^K s_k$ and $\tilde{c}=\sum_{k=1}^K \tilde{s}_k$ such that  $s_k\in \mathcal{X}_k$ and $\tilde{c}, c\in \mathcal{Q} = \uplus_{k =1}^K \mathcal{X}_k$.
%By the Cauchy-Schwarz inequality, we have $a b>|c|^2$ and $a\tilde{b}>|\tilde{c}|^2$, and then
%\begin{align*}
%&\det\Big\{(\mathbf{X}_2^H \mathbf{D}\mathbf{X}_2 +\sigma^2 \mathbf{I}_2)(\widetilde{\mathbf X}_2^H \mathbf{D}\widetilde{\mathbf X}_2 +\sigma^2 \mathbf{I}_2)^{-1}) \Big\}%=\det(\mathbf{C})/\det(\mathbf{C}')
%=\frac{ab-|c|^2}{a \tilde{b}-|\tilde{c}|^2}>0,\\
%&{\rm tr}\Big\{(\mathbf{X}_2^H \mathbf{D}\mathbf{X}_2 +\sigma^2 \mathbf{I}_2)(\widetilde{\mathbf X}_2^H \mathbf{D}\widetilde{\mathbf X}_2 +\sigma^2 \mathbf{I}_2)^{-1}\Big\}\\
%&=\frac{1}{a\tilde{b}-|\tilde{c}|^2}{\rm tr}\Bigg (\begin{bmatrix}
%a & c\\
%c^* & b
%\end{bmatrix}
%\begin{bmatrix}
%\tilde{b} & -\tilde{c}\\
%-\tilde{c}^* & a
%\end{bmatrix}\Bigg )
%=\frac{1}{a\tilde{b}-|\tilde{c}|^2}{\rm tr}\Bigg (\begin{bmatrix}
%a\tilde{b}-c \tilde{c}^* &  ac-a\tilde{c}\\
%\tilde{b}c^*-b \tilde{c}^* & ab -c^*\tilde{c}
%\end{bmatrix}\Bigg )\\
%&=\frac{ab+a\tilde{b} -c\tilde{c}^*- c^* \tilde{c}}{a\tilde{b}-|\tilde{c}|^2}>0.
%\end{align*}
%where $s_k\in \mathcal{X}_k$ and $(\sum_{k=1}^K s_k), (\sum_{k=1}^K \tilde{s}_k)\in \mathcal{Q} = \uplus_{k =1}^K \mathcal{X}_k$.

We can see from~\eqref{eqn:correlationmatrix} that $\mathbf{X}_2^H \mathbf{D}\mathbf{X}_2 +\sigma^2 \mathbf{I}_2$ and $\widetilde{\mathbf{X}}_2^H \mathbf{D}\widetilde{\mathbf{X}}_2 +\sigma^2 \mathbf{I}_2$ are independent of the permutation function $\pi$, but depends on the power allocation vector $\mathbf{p}=[p_1, \ldots, p_K]^T$, and the information carrying vectors $\mathbf{s}=[s_1, \ldots, s_K]^T$ and $\tilde{\mathbf{s}}=[\tilde{s}_1, \ldots, \tilde{s}_K]^T$.
In this case, the ML receiver given in~\eqref{eqn:simplifiedMLreceiver} can be further simplified as
\begin{align}
\widehat{\mathbf{X}}_2&={\arg\min}_{\mathbf{X}_2}~\frac{ a \|\mathbf{y}_2\|^2 + b \|\mathbf{y}_1\|^2 -2 \Re( c\mathbf{y}_1^H \mathbf{y}_2)}{ab-|c|^2}\nonumber + M \ln \big(ab-|c|^2\big),
\end{align}
Inserting~\eqref{eqn:correlationmatrix} into~\eqref{eqn:KLdistanece}, and after some algebraic manipulations, we have
%\begin{subequations}
%	\begin{align*}
%	&\mathcal{D}_{\rm KL}(\mathbf{X}_2 ||\widetilde{\mathbf X}_2)= \frac{ab+a\tilde{b} -c\tilde{c}^*- c^* \tilde{c}}{a\tilde{b}-|\tilde{c}|^2} - \ln\Big (\frac{ab-|c|^2}{a\tilde{b}-|\tilde{c}|^2} \Big )-2\\
%	&= \frac{ab+a\tilde{b}-|c|^2 -|\tilde{c}|^2+ |c|^2+|\tilde{c}|^2-c\tilde{c}^*- c^* \tilde{c}}{a\tilde{b}-|\tilde{c}|^2} - \ln\Big (\frac{ab-|c|^2}{a\tilde{b}-|\tilde{c}|^2}\Big ) -2\\
%	%&=\frac{ab-|c|^2}{a\tilde{b}-|\tilde{c}|^2}+\frac{|c|^2+|\tilde{c}|^2-c\tilde{c}^*- c^* \tilde{c}}{a\tilde{b}-|\tilde{c}|^2} - \ln\Big (\frac{ab-|c|^2}{a\tilde{b}-|\tilde{c}|^2}\Big )-1\\
%	%&=\frac{ab-|c|^2}{a\tilde{b}-|\tilde{c}|^2} - \ln\Big (\frac{ab-|c|^2}{a\tilde{b}-|\tilde{c}|^2}\Big )-1+\frac{|c-\tilde{c}|^2}{a\tilde{b}-|\tilde{c}|^2}.
%	&=\underbrace{\frac{ab-|c|^2}{a\tilde{b}-|\tilde{c}|^2}-\ln\Big (\frac{ab-|c|^2}{a\tilde{b}-|\tilde{c}|^2}\Big )-1}_{\rm T_1}+\underbrace{\frac{|c-\tilde{c}|^2}{a\tilde{b}-|\tilde{c}|^2}}_{\rm T_2}.
%	\end{align*}
%\end{subequations}
\begin{subequations}
\begin{align*}
&\mathcal{D}_{\rm KL}(\mathbf{X}_2 ||\widetilde{\mathbf X}_2)=
f_1(\mathbf{p},\mathbf{s}, \tilde{\mathbf{s}})+f_2(\mathbf{p},\mathbf{s}, \tilde{\mathbf{s}}),
\end{align*}
\end{subequations}
where
\begin{align*}
&f_1(\mathbf{p},\mathbf{s}, \tilde{\mathbf{s}})=\frac{\big (\sum_{k=1}^K \frac{1}{p_k}+\sigma^2\big ) \big (\sum_{k=1}^K p_k |s_k|^2 +\sigma^2 \big )-\big|\sum_{k=1}^K s_k\big|^2}{\big (\sum_{k=1}^K \frac{1}{p_k}+\sigma^2\big )\big (\sum_{k=1}^K p_k |\tilde{s}_k|^2 +\sigma^2\big )-\big|\sum_{k=1}^K \tilde{s}_k\big|^2}\\
&\qquad\qquad\quad-\ln \left[ \frac{\big (\sum_{k=1}^K \frac{1}{p_k}+\sigma^2\big )\big (\sum_{k=1}^K p_k |s_k|^2 +\sigma^2\big )-\big|\sum_{k=1}^K s_k\big|^2}{\big (\sum_{k=1}^K \frac{1}{p_k}+\sigma^2\big )\big (\sum_{k=1}^K p_k |\tilde{s}_k|^2 +\sigma^2\big )-\big|\sum_{k=1}^K \tilde{s}_k\big|^2}\right]-1,\\
&f_2(\mathbf{p},\mathbf{s}, \tilde{\mathbf{s}})=\frac{\big|\sum_{k=1}^K s_k-\sum_{k=1}^K \tilde{s}_k\big|^2}{\big (\sum_{k=1}^K \frac{1}{p_k}+\sigma^2\big )\big (\sum_{k=1}^K p_k |\tilde{s}_k|^2 +\sigma^2\big )-\big|\sum_{k=1}^K \tilde{s}_k\big|^2}.
\end{align*}
Recall that the power constraints are $\mathbb E\{|x_{k,t}|^2\} \le P_k$ for $k=1,\ldots, K$ and $t=1,2$. That is, for the first and second time slots, we have
$\mathbb E\{|x_{k,1}|^2\}=\frac{1}{p_{\pi(k)}\beta_{k}}  \le P_k$, and $\mathbb E\{|x_{k,2}|^2\}=\frac{p_{\pi(k)} E_{\pi(k)}d^2}{\beta_k}  \le P_k$, where $E_{k} =\frac{\mathbb E\{|s_{k}|^2\}}{d^2}$. The power constraints can thus be expressed as:
\begin{align}
\frac{1}{P_{\pi^{-1}(k)} \beta_{\pi^{-1}(k)}}\le p_k \le \frac{P_{\pi^{-1}(k)} \beta_{\pi^{-1}(k)}}{E_k d^2}, \quad k=1,\ldots, K.
\end{align}

Our design can now be formulated into the following optimization problem.
\begin{problem}\label{problem1} Find the optimal power control vector $\mathbf{p}$ and permutation $\pi$ under individual average power constraints:
\begin{subequations}\label{eqn:problem1}
\begin{align}
&\max_{\{\pi,\mathbf{p}\}}\,\min_{\{\mathbf{s}, \tilde{\mathbf{s}}:\,\mathbf{s}\neq \tilde{\mathbf{s}}\}} ~f_1(\mathbf{p},\mathbf{s}, \tilde{\mathbf{s}})+f_2(\mathbf{p},\mathbf{s}, \tilde{\mathbf{s}})\\
&{\rm ~s.t.},~\frac{1}{P_{\pi^{-1}(k)} \beta_{\pi^{-1}(k)}}\le p_k \le \frac{P_{\pi^{-1}(k)} \beta_{\pi^{-1}(k)}}{E_k d^2}, \quad k=1,\ldots, K.
\end{align}
\end{subequations}
\hfill\QED
\end{problem}

For Problem~\ref{problem1}, we first can attain that $f_1(\mathbf{p},\mathbf{s}, \tilde{\mathbf{s}}) \ge 0$ by applying the fundamental inequality in information theory~\cite[Lemma\,2.29]{Yeu08network}, where the equality $f_1(\mathbf{p},\mathbf{s}, \tilde{\mathbf{s}})=0$ holds if and only if
\begin{align}\label{eqn:optcondition}
\Big(\sum_{k=1}^K \frac{1}{p_k}+\sigma^2\Big)\Big(\sum_{k=1}^K p_k (|s_k|^2- |\tilde{s}_k|^2) \Big)-\Big(\big|\sum_{k=1}^K s_k\big|^2-\big|\sum_{k=1}^K \tilde{s}_k\big|^2\Big)=0.
\end{align}

Considering the fact that the joint minimization of $f_1(\mathbf{p},\mathbf{s}, \tilde{\mathbf{s}})$ and $f_2(\mathbf{p},\mathbf{s}, \tilde{\mathbf{s}})$ over $\{\mathbf{s}, \tilde{\mathbf{s}}:\mathbf{s}\neq \tilde{\mathbf{s}}\}$ could be extremely tedious, we consider the minimization of $f_2(\mathbf{p},\mathbf{s}, \tilde{\mathbf{s}})$ first, which is a lower bound of $\mathcal{D}_{\rm KL}(\mathbf{X}_2 ||\widetilde{\mathbf X}_2)$ as $f_1(\mathbf{p},\mathbf{s}, \tilde{\mathbf{s}})\ge 0$. We will verify the condition when the minimum of $f_1(\mathbf{p},\mathbf{s}, \tilde{\mathbf{s}})$ and $f_2(\mathbf{p},\mathbf{s}, \tilde{\mathbf{s}})$ can be achieved simultaneously. Mathematically, we temporarily focus on solving the following optimization problem:
\begin{problem}\label{pbm:problem2} Find the power control coefficients $\mathbf{p}$ and permutation $\pi$, such that
\begin{subequations}\label{eqn:equalpowerdesigngeneral}
\begin{align}
&\max_{\{\pi, \mathbf{p}\}}\,\min_{\{\mathbf{s}, \tilde{\mathbf{s}}:\,\mathbf{s}\neq \tilde{\mathbf{s}}\}}~f_2(\mathbf{p},\mathbf{s}, \tilde{\mathbf{s}})=\frac{\big|\sum_{k=1}^K s_k-\sum_{k=1}^K \tilde{s}_k\big|^2}{(\sum_{k=1}^K \frac{1}{p_k}+\sigma^2)(\sum_{k=1}^K p_k |\tilde{s}_k|^2 +\sigma^2)-|\sum_{k=1}^K \tilde{s}_k|^2}\label{eqn:f2objec}\\
&~{\rm s.t.}~\frac{1}{P_{\pi^{-1}(k)} \beta_{\pi^{-1}(k)}}\le p_k \le \frac{P_{\pi^{-1}(k)} \beta_{\pi^{-1}(k)}}{E_k d^2}, \quad k=1,\ldots, K.\label{eqn:f2constraint}
\end{align}
\end{subequations}
\hfill\QED
\end{problem}
%The solution can be summarized into the following lemma.
%\begin{lemma}\label{lem:optconstellation}
%The optimal solution for $\tilde{\mathbf{s}}=[\tilde{s}_1, \ldots, \tilde{s}_K]^T$, where $\tilde{s}_k=(\hat{v}_k+j \hat{w}_k)\}_{k=1}^K$, such that
%	\begin{align*}
%	\hat v_k =\begin{cases}\big(2^{N_{I,k} -1}-\frac{1}{2}\big) \times 2^{\sum_{\ell=1}^{k-1} N_{I,\ell} }, &{\rm for~}k=1,\ldots, K-1;\\
%	-\big(2^{N_{I,K} -1}-\frac{1}{2}\big) \times 2^{\sum_{\ell=1}^{K-1} N_{I,\ell} }, &{\rm for~} k=K.
%	\end{cases}
%	\end{align*}
%	\begin{align*}
%	\hat w_k =\begin{cases}\big(2^{N_{Q,k} -1}-\frac{1}{2}\big) \times 2^{\sum_{\ell=1}^{k-1} N_{Q,\ell} }, &{\rm for~}k=1,\ldots, K-1;\\
%	-\big(2^{N_{Q,K} -1}-\frac{1}{2}\big) \times 2^{\sum_{\ell=1}^{K-1} N_{Q,\ell} }, &{\rm for~} k=K.
%	\end{cases}
%	\end{align*}
%	\hfill\QED
%\end{lemma}

We first consider the inner optimization problem in Problem~\ref{pbm:problem2}. The denominator of~\eqref{eqn:f2objec} is independent of $\mathbf{s}$ and the numerator is minimized when the sum terms $\sum_{k=1}^K s_k$ and $\sum_{k=1}^K \tilde{s}_k$ are the neighboring points on the sum-constellation, where the minimum value of $\big|\sum_{k=1}^K s_k-\sum_{k=1}^K \tilde{s}_k\big|^2$ is $d^2$.  For notation simplicity, we define $\tilde{\mathbf{s}}=(\tilde{\mathbf{v}}+j\tilde{\mathbf{w}})d$, where  $\tilde{\mathbf{v}}=[\tilde{v}_1, \ldots, \tilde{v}_K]^T$ and $\tilde{\mathbf{w}}=[\tilde{w}_1, \ldots, \tilde{w}_K]^T$. As the power constraint given in~\eqref{eqn:f2constraint} is independent of $\mathbf{v}$ and $\mathbf{w}$, Problem~\ref{pbm:problem2} can be split into two subproblems:
\begin{subequations}\label{eqn:optinphase}
\begin{align}
&\max_{\tilde{\mathbf{v}}}~f_3(\tilde{\mathbf{v}})=\Big (\sum_{k=1}^K \frac{1}{p_k}+\sigma^2\Big )\Big (\sum_{k=1}^K p_k \tilde{v}_k^2  +\frac{\sigma^2}{d^2}\Big )- \Big (\sum_{k=1}^K \tilde{v}_k\Big )^2\\
&{\rm ~s.t.}~\tilde{v}_k \in
\Big\{\pm \Big (m-\frac{1}{2}\Big ) \times 2^{\sum_{\ell=1}^{k-1} N_{I,\ell} }\Big\}_{m=1}^{2^{N_{I,k}-1}},\quad k=1,\ldots, K.
\end{align}
\end{subequations}
and
\begin{subequations}\label{eqn:optquadrature}
\begin{align}
&\max_{\tilde{\mathbf{w}}}~f_4(\tilde{\mathbf{w}})=\Big (\sum_{k=1}^K \frac{1}{p_k}+\sigma^2\Big )\Big (\sum_{k=1}^K p_k \tilde{w}_k^2  +\frac{\sigma^2}{d^2}\Big )- \Big (\sum_{k=1}^K \tilde{w}_k\big )^2\\
&{\rm ~s.t.}~\tilde{w}_k \in \Big\{\pm \Big (m-\frac{1}{2}\Big ) \times 2^{\sum_{\ell=1}^{k-1} N_{Q,\ell} }\Big\}_{m=1}^{2^{N_{Q,k}-1}},\quad k=1,\ldots, K.
\end{align}
\end{subequations}

In the following, we only present the maximization of $f_3(\tilde{\mathbf{v}})$ over $\tilde{\mathbf{v}}$ in~\eqref{eqn:optinphase}, and the maximization of $f_4(\tilde{\mathbf{w}})$ over $\tilde{\mathbf{w}}$ given in ~\eqref{eqn:optquadrature} follows similarly and hence is omitted fore brevity. We now rewrite the objective function~(\ref{eqn:optinphase}a) as
\begin{subequations}
	\begin{align*}
	f_3(\tilde{\mathbf{v}})=&\bigg (\frac{1}{p_1}+\Big (\sum_{k=2}^K \frac{1}{p_k} +\sigma^2 \Big )\bigg )\Big (p_1 \tilde{v}_1^2 +\big (\sum_{k=2}^K p_k \tilde{v}_k^2 +\frac{\sigma^2}{d^2}\big )\Big )- \Big (\tilde{v}_1+\sum_{k=2}^K \tilde{v}_k\Big )^2\\
	=&\frac{1}{p_1}\Big (\sum_{k=2}^K p_k \tilde{v}_k^2  +\frac{\sigma^2}{d^2}\Big )+ p_1 \tilde{v}_1^2\Big (\sum_{k=2}^K \frac{1}{p_k}+\sigma^2\Big )+\Big (\sum_{k=2}^K \frac{1}{p_k}+\sigma^2\Big )\Big (\sum_{k=2}^K p_k \tilde{v}_k^2  +\frac{\sigma^2}{d^2}\Big )\\
	& -2 \tilde{v}_1\big (\sum_{k=2}^K \tilde{v}_k\big )-\big (\sum_{k=2}^K \tilde{v}_k\big )^2\\
	=&\frac{1}{p_1}\Big (\sum_{k=2}^K p_k \tilde{v}_k^2  +\frac{\sigma^2}{d^2}\Big ) +\frac{1}{p_2}\Big (\sum_{k=3}^K p_k \tilde{v}_k^2 +\frac{\sigma^2}{d^2}\Big )+ p_1 \tilde{v}_1^2\Big (\sum_{k=2}^K \frac{1}{p_k}+\sigma^2\Big )+ p_2 \tilde{v}_2^2\Big (\sum_{k=3}^K \frac{1}{p_k}+\sigma^2\Big )\\
	&+\Big (\sum_{k=3}^K \frac{1}{p_k}+\sigma^2\Big )\Big (\sum_{k=3}^K p_k \tilde{v}_k^2  +\frac{\sigma^2}{d^2}\Big)-\big(\sum_{k=3}^K \tilde{v}_k\big)^2-2 \tilde{v}_1\big(\sum_{k=2}^K \tilde{v}_k\big)-2 \tilde{v}_2\big(\sum_{k=3}^K \tilde{v}_k\big)\\
	%&\qquad \vdots\\
	=&\sum_{\ell=1}^{K-1}\frac{1}{p_\ell} \Big(\sum_{k=\ell+1}^K p_k \tilde{v}_k^2 +\frac{\sigma^2}{d^2}\Big)+\sum_{\ell=1}^{K-1}p_\ell \tilde{v}_\ell^2 \Big(\sum_{k=\ell+1}^K \frac{1}{p_k} +\sigma^2\Big) +p_K \tilde{v}_K^2\sigma^2  \\
	&+\frac{\sigma^4}{d^2}+p_K \tilde{v}_K^2\sigma^2  +\frac{\sigma^4}{d^2}-2\sum_{\ell=1}^{K-1} \tilde{v}_\ell\sum_{k=\ell+1}^K \tilde{v}_k\\
	=&f_5(\tilde{\mathbf{v}})-f_6(\tilde{\mathbf{v}}),
	\end{align*}
\end{subequations}
where $f_5(\tilde{\mathbf{v}})=\sum_{\ell=1}^{K-1}\frac{1}{p_\ell} \Big(\sum_{k=\ell+1}^K p_k \tilde{v}_k^2 +\frac{\sigma^2}{d^2}\Big)+\sum_{\ell=1}^{K-1}p_\ell \tilde{v}_\ell^2 \Big(\sum_{k=\ell+1}^K \frac{1}{p_k} +\sigma^2\Big) +p_K \tilde{v}_K^2\sigma^2  +\frac{\sigma^4}{d^2}+p_K \tilde{v}_K^2\sigma^2  +\frac{\sigma^4}{d^2}$, and $f_6(\tilde{\mathbf{v}})=2\sum_{\ell=1}^{K-1} \tilde{v}_\ell\sum_{k=\ell+1}^K \tilde{v}_k$.
We then can maximize $f_5(\tilde{\mathbf{v}})-f_6(\tilde{\mathbf{v}})$. In what follows, we will show that the maximization of  $f_5(\tilde{\mathbf{v}})$ and the minimization of $f_6(\tilde{\mathbf{v}})$  can be achieved simultaneously. First, we can observe that the maximization of $f_5(\tilde{\mathbf{v}})$ is achieved when $|\tilde{v}_k|$, $k=1,\ldots, K$, are maximized for signal transmitted from every user. We next consider the minimization of $f_6(\tilde{\mathbf{v}})$.  To that end, we have,
\begin{align}
\frac{\partial f_6(\tilde{\mathbf{v}})}{\partial \tilde{v}_k} =2\sum_{\ell=1, \ell\neq k}^K \tilde{v}_\ell, k=1, \ldots, K.
\end{align}
The optimal value can be attained by enumeration of $\tilde{v}_K \in \Big\{\big(m-\frac{1}{2}\big) \times 2^{\sum_{\ell=1}^{K-1} N_\ell }\Big\}_{m=1}^{2^{N_K-1}}$, and $\tilde{v}_K \in \Big\{-\big(m-\frac{1}{2}\big) \times 2^{\sum_{\ell=1}^{K-1} N_\ell }\Big\}_{m=1}^{2^{N_K-1}}$.
\begin{enumerate}
\item If $\tilde{v}_K \in
\Big\{\big(m-\frac{1}{2}\big) \times 2^{\sum_{\ell=1}^{K-1} N_\ell }\Big\}_{m=1}^{2^{N_K-1}}$, then for any $\tilde{v}_k \in
\Big\{\pm \big(m-\frac{1}{2}\big) \times 2^{\sum_{\ell=1}^{k-1} N_\ell }\Big\}_{m=1}^{2^{N_k-1}}$, $k=1,\ldots, K-1$, we have
\begin{align*}
&\frac{\partial f_6(\tilde{\mathbf{v}})}{\partial \tilde{v}_k}
=2 \tilde{v}_K+2\sum_{\ell=1, \ell\neq k}^{K-1} \tilde{v}_\ell\ge 2 \min \tilde{v}_K+ 2\min_{\{\tilde{v}_\ell\}_{\ell=1}^{K-1}} \sum_{\ell=1, \ell\neq k}^{K-1} \tilde{v}_\ell\\
&> 2^{\sum_{\ell=1}^{K-1} N_\ell } + 2\min_{\{ \tilde{v}_\ell\}_{\ell=1}^{K-1}} \sum_{\ell=1}^{K-1} \tilde{v}_\ell= 2^{\sum_{\ell=1}^{K-1} N_\ell} - 2  \Big(2^{\sum_{\ell=1}^{K-1} N_\ell-1}-\frac{1}{2}\Big)=1.
\end{align*}
In this case, the optimal value of $\{\tilde{v}_k\}_{k=1}^{K-1}$ to minimize $f_6(\tilde{\mathbf{v}})$ is given by
\begin{align*}
\tilde{v}_k= -\big(2^{N_k}-1\big) \times 2^{\sum_{\ell=1}^{k-1} N_\ell -1}, {\rm~for~}k=1, \ldots, K-1.
\end{align*}
Note that $\frac{\partial f_6(\tilde{\mathbf{v}})}{\partial \tilde{v}_K}=2\sum_{\ell=1}^{K-1} \tilde{v}_\ell<0$, then for $\tilde{v}_K \in
\Big\{\big(m-\frac{1}{2}\big) \times 2^{\sum_{\ell=1}^{K-1} N_\ell }\Big\}_{m=1}^{2^{N_K-1}}$, the optimal value of $\tilde{v}_K$ is $\tilde{v}_K=\big(2^{N_K}-1\big) \times 2^{\sum_{\ell=1}^{K-1} N_\ell -1}$.
\item If  $\tilde{v}_K \in \Big\{-\big(m-\frac{1}{2}\big) \times 2^{\sum_{\ell=1}^{K-1} N_\ell }\Big\}_{m=1}^{2^{N_K-1}}$, for $\tilde{v}_k \in
\Big\{\pm \big(m-\frac{1}{2}\big) \times 2^{\sum_{\ell=1}^{k-1} N_\ell }\Big\}_{m=1}^{2^{N_k-1}}$, $k=1,\ldots, K-1$, we have
\begin{align*}
&\frac{\partial f_6(\tilde{\mathbf{v}})}{\partial \tilde{v}_k}
=2 \tilde{v}_K+2\sum_{\ell=1, \ell\neq k}^{K-1} \tilde{v}_\ell
\le 2 \tilde{v}_K+ 2\max_{\{ \tilde{v}_\ell\}_{\ell=1}^{K-1}} \sum_{\ell=1, \ell\neq k}^K \tilde{v}_\ell\\
&< -2^{\sum_{\ell=1}^{K-1} N_\ell }+ 2\max_{\{\tilde{v}_\ell\}_{\ell=1}^{K-1}} \sum_{\ell=1}^{K-1} \tilde{v}_\ell= -2^{\sum_{\ell=1}^{K-1} N_\ell }+ 2\Big( 2^{\sum_{\ell=1}^{K-1} N_\ell-1}-\frac{1}{2}\Big)=-1.
 \end{align*}
In this case, the optimal value of $\{\tilde{v}_k\}_{k=1}^{K-1}$ to minimize $f_6(\tilde{\mathbf{v}})$ is given by
\begin{align*}
\tilde{v}_k= \Big(2^{N_k}-\frac{1}{2}\Big) \times 2^{\sum_{\ell=1}^{k-1} N_\ell }, {\rm~for~}k=1, \ldots, K-1.
\end{align*}
In addition, we note that $\frac{\partial f_6(\tilde{\mathbf{v}})}{\partial \tilde{v}_K}=2\sum_{\ell=1}^{K-1} \tilde{v}_\ell<0$, then for  $\tilde{v}_K \in
\Big\{-\big(m-\frac{1}{2}\big) \times 2^{\sum_{\ell=1}^{K-1} N_\ell }\Big\}_{m=1}^{2^{N_K-1}}$, the optimal value of $\tilde{v}_K$ is $\tilde{v}_K=-\big(2^{N_K}-1\big) \times 2^{\sum_{\ell=1}^{K-1} N_\ell -1}$.
\end{enumerate}

Overall, the maximum value of $f_6(\tilde{\mathbf{v}})=2\sum_{\ell=1}^{K-1} \tilde{v}_\ell\sum_{k=\ell+1}^K \tilde{v}_k$ can be achieved by $\tilde{\mathbf{v}}^\star=[\tilde{v}_1^\star, \ldots, \tilde{v}_K^\star]^T$ where
\begin{align}\label{eqn:optcase1}
\tilde{v}_k^\star =\begin{cases}-\big(2^{N_{I,k} -1}-\frac{1}{2}\big) \times 2^{\sum_{\ell=1}^{k-1} N_{I,\ell} }, &{\rm for~}k=1,\ldots, K-1;\\
\big(2^{N_{I,K} -1}-\frac{1}{2}\big) \times 2^{\sum_{\ell=1}^{K-1} N_{I,\ell} }, &{\rm for~} k=K,
\end{cases}
\end{align}
or equivalently
\begin{align}\label{eqn:optcase2}
\tilde{v}_k^\star =\begin{cases}\big(2^{N_{I,k} -1}-\frac{1}{2}\big) \times 2^{\sum_{\ell=1}^{k-1} N_{I,\ell} }, &{\rm for~}k=1,\ldots, K-1;\\
-\big(2^{N_{I,K} -1}-\frac{1}{2}\big) \times 2^{\sum_{\ell=1}^{K-1} N_{I,\ell} }, &{\rm for~} k=K.
\end{cases}
\end{align}

For both cases, we can observe that $f_5(\tilde{\mathbf{v}})$ is also maximized by $\tilde{\mathbf{v}}^\star$. Due to the symmetry of the solutions given in~\eqref{eqn:optcase1} and~\eqref{eqn:optcase2}, in what follows, we only consider the solution given in~\eqref{eqn:optcase1}. In this case, the sum-constellation for achieving the inner minimum is
\begin{align}\label{eqn:optsumcons}
&\sum_{k=1}^K \tilde{s}_k=\sum_{k=1}^K\tilde{v}_k+j \tilde{w}_k\nonumber\\
&=\bigg [\frac{1+j}{2}+\big (2^{N_{I,K} -1}-1\big ) \times 2^{\sum_{\ell=1}^{K-1} N_{I,\ell} }+j\big (2^{N_{Q,K} -1}-1\big ) \times 2^{\sum_{\ell=1}^{K-1} N_{Q,\ell} }\bigg ] d.
\end{align}

We then can have the following remark:
\begin{remark}
When $N_{I,K}=N_{Q,K}=1$, the solution given in~\eqref{eqn:optcase1}, which minimizes $f_2(\mathbf{p},\mathbf{s}, \tilde{\mathbf{s}})$ also minimizes $f_1(\mathbf{p},\mathbf{s}, \tilde{\mathbf{s}})$. \hfill$\Box$
\end{remark}

\emph{Proof:} For the solution of $\tilde{\mathbf{s}}$ given in~\eqref{eqn:optcase1}, the sum constellation is given in~\eqref{eqn:optsumcons}. When $N_{I,K}=N_{Q,K}=1$, we have $\sum_{k=1}^K \tilde{s}_k=\frac{1+j}{2}d$, and we can let $\sum_{k=1}^K s_k=\frac{1-j}{2}d$.  Inserting them back into~\eqref{eqn:optcondition}, we have
$f_1(\mathbf{p},\mathbf{s}, \tilde{\mathbf{s}})=0$. That is, the values that minimize $f_2(\mathbf{p},\mathbf{s}, \tilde{\mathbf{s}})$ also minimizes $f_1(\mathbf{p},\mathbf{s}, \tilde{\mathbf{s}})$.
\hfill$\Box$

We now consider the outer optimization problem, where the objective function is a monotonically decreasing function against the term $\frac{ab}{d^2}$.
%We note that $\frac{ab}{d^2}= \frac{(\sum_{k=1}^K \frac{1}{p_k}+\sigma^2)(\sum_{k=1}^K p_k E_k d^2 +\sigma^2)}{d^2}$, where $E_k=v_k^2+w_k^2$ and hence Problem~\ref{pbm:problem2} can be reformulated by:
The optimization problem can be reformulated as
\begin{subequations}\label{eqn:poweropt1}
	\begin{align}
	&\max_{\pi, \mathbf{p}}~\bigg (\sum_{k=1}^K \frac{1}{p_k}+\sigma^2\bigg )\bigg (\sum_{k=1}^K p_k E_k  +\frac{\sigma^2}{d^2}\bigg )\label{eqn18:objectivefun}\\
	&\quad {\rm s.t.}~
	\frac{1}{p_k}\le P_{\pi^{-1}(k)} \beta_{\pi^{-1}(k)},~ p_k E_k d^2\le P_{\pi^{-1}(k)} \beta_{\pi^{-1}(k)}, \quad k=1,\ldots, K.\label{eqn18:const3}
	\end{align}
\end{subequations}
The optimization problem in~\eqref{eqn:poweropt1} can be resolved by first fixing $\pi$ to find the optimal value of $\mathbf{p}$, and then perform further optimization on $\pi$. To that end, we can observe from~\eqref{eqn18:const3} that, for any given $\pi$, the feasible range of $d^2$ is given by $d^2 \le \frac{P_{\pi^{-1}(k)} \beta_{\pi^{-1}(k)}}{p_k E_k} \le \frac{P_{\pi^{-1}(k)}^2 \beta_{\pi^{-1}(k)}^2}{E_k}$ for $k=1,\ldots, K$, or equivalently $d^2 \le \min\,\Big\{\frac{P_{\pi^{-1}(k)}^2 \beta_{\pi^{-1}(k)}^2}{E_k}\Big\}_{k=1}^K$.
By the Cauchy-Swartz inequality, we have
\begin{align*}
&\bigg (\sum_{k=1}^K \frac{1}{p_k}+\sigma^2\bigg )\bigg (\sum_{k=1}^K p_k E_k  +\frac{\sigma^2}{d^2}\bigg )\\
&\stackrel{(a)}{\ge} \bigg (\sum_{k=1}^K \frac{1}{\sqrt{p_k}} \sqrt{p_k E_k d^2} +\frac{\sigma^2}{d}\bigg )^2=\bigg (\sum_{k=1}^K  \sqrt{E_k} +\frac{\sigma^2}{d}\bigg )^2,
\end{align*}
where the inequality in~$(a)$ holds if and only if $\frac{\sqrt{p_k E_k}}{{1}/{\sqrt{p_k}}}=\frac{1}{d}$, for $k=1,\ldots K$. Or equivalently, the optimal power allocation is $\mathbf{p}=[p_1^\star, \ldots, p^\star]^T$ where $p_k^\star=\frac{1}{\sqrt{E_k} d}$ for $k=1, \ldots, K$. Our next task is to check the power constraint on $p_k^\star$ given in~\eqref{eqn18:const3} is violated or not. For $d^2 \le \min\,\Big\{\frac{P_{\pi^{-1}(k)}^2 \beta_{\pi^{-1}(k)}^2}{E_k}\Big\}_{k=1}^K$, we have
\begin{align}
&\frac{1}{p_k^\star}=\sqrt{E_k}d \le P_{\pi^{-1}(k)} \beta_{\pi^{-1}(k)},\quad p_k^\star E_k d^2=\sqrt{E_k} d \le P_{\pi^{-1}(k)} \beta_{\pi^{-1}(k)}, {\rm~for~}k=1,\ldots, K,
\end{align}
where no power constraints are violated for $\mathbf{p}$. Finally, the optimization problem on $\pi$ can be given by
\begin{align}
&\min_{\pi}~\sum_{k=1}^{K} \sqrt{E_k}+\frac{\sigma^2}{d}
\quad {\rm s.t.}~d^2\le \frac{P_{\pi^{-1}(k)}^2 \beta_{\pi^{-1}(k)}^2}{E_k}, \quad k=1, \ldots, K.
\end{align}
Or equivalently, we aim to solve
\begin{align}
&\max_{\pi}~d
\quad {\rm s.t.}~d^2 \le \frac{P_k^2 \beta_k^2}{E_{\pi(k)}} , \quad k=1, \ldots, K.
\end{align}
Before proceeding on, we establish the following lemma.
\begin{lemma}\label{lemma:orderedseq}
	Suppose that two positive sequences $\{a_n\}_{n=1}^N$ and $\{b_n\}_{n=1}^N$ are arranged both in a nondecreasing order. If we let	$\Pi$ denote the set containing all the possible permutations of $1,2,\cdots, N$, then, the solution to
	%in which $(\pi(k), k=1,2,\ldots, N)$ is a permutation on $(1,2, \cdots, N)$ and $\pi(k)$ is the $k$-th entry of $(\pi(k), k=1,2,\ldots, N)$.
	the optimization problem,
	%\begin{align}
	$\max_{\pi \in \Pi} \min \Big\{ \frac{a_k}{b_{\pi(k)}}\Big\}_{k=1}^K$,
	%\end{align}
	is given by $\pi^\star(k)=k$ for $k=1,2,\cdots, K$.~\hfill\QED
\end{lemma}

By Lemma~\ref{lemma:orderedseq}, and note that $P_1\beta_1 \le \ldots \le P_k\beta_K$, to maximize $d$, we should let $E_{\pi(1)} \le \ldots \le E_{\pi(K)}$, i.e., the average power of the sub-constellations should be arranged in an ascending order. All the above discussions can be summarized into the following theorem:
\begin{theorem}
	The users are ordered such that $P_1\beta_1 \le P_2\beta_2 \le \ldots \le P_k\beta_K$, and we define $d^\star =\min_{k} \big\{\frac{P_k\beta_k}{\sqrt{E_k}}\big\}_{k=1}^K$, the optimal transmit power for all users can be given by $\mathbf{p}^\star=[\frac{1}{\sqrt{E_1} d^\star}, \ldots, \frac{1}{\sqrt{E_K} d^\star}]^T$. And the optimal permutation matrix is the identity matrix, i.e., $\mathbf{\Pi}=\mathbf{I}_K$.
	\hfill\QED
\end{theorem}

%\section{Calculation of Simulations}
%We assume that $\mathcal{X}_1 = \big\{\pm (m-\frac{1}{2})d \pm j(n-\frac{1}{2})d\big\}_{m,n=1}^{\sqrt{K}/2}$,
%and
%$\mathcal{X}_k =
%\big\{\big(\pm (m-\frac{1}{2}) \pm j(n-\frac{1}{2})\big)d \times \sqrt{K}^{(k-1)/2}d\big\}_{m=1}^{\sqrt{K}/2}$  for $k \ge 2$.

\section{Simulation Results and Discussions}
In this section, computer simulations are performed to demonstrate the superior performance of our design in comparison with existing benchmarks. In our simulations, the small-scale fading is assumed to be the normalized Rayleigh fading. The path-loss as a function of transmission distance $d$ at antenna far-field can be approximated by
\begin{align*}
10\log_{10} L = 20\log_{10} \Big(\frac{\lambda}{4\pi d_0}\Big ) -10\gamma \log_{10} \Big(\frac{d}{d_0}\Big) - \psi,\quad d\ge d_0,
\end{align*}
where $d_0=100$m is the reference distance, $\lambda=v_c/f_c$ ($f_c=3$GHz) is the wavelength of carrier, $\gamma=3.71$ is the path-loss exponent~\cite{goldsmith05}. In the above model, $\psi \sim \mathcal{N}(0, \sigma_{\psi}^2)$ ($\sigma_{\psi} =3.16$) is the Gaussian random shadowing attenuation resulting from the blockage of objects. For the receiver, we assume that the noise power is %$\sigma_n^2 = N_0 B = 1.38\times 10^{-23}\times 290\times 10^{6/10}\times 20\times 10^6= 3.2\times 10^{-13} \,{\rm W}$,
$10\log_{10}\sigma^2 =10\log_{10} N_0 B_w = 10\log_{10} 3.2\times 10^{-10} = -125.97\,{\rm dB}$ where the channel bandwidth $B_w=20$MHz, and $N_0= k_0 T_0 10^{F_0/10}$ is the power spectral density of noise with $k_0=1.38\times10^{-23}$ J/K being the Boltzman's constant, reference temperature $T_0 =290$K (``room temperature''), and noise figure $F_0=6$\,dB. For clarity, all the simulation parameters are summarized in Table~\ref{tbl:simpar}.
\begin{table}[ht]
	\caption{Simulation Parameters} % title of Table
	\begin{center}
		\begin{tabular}{|c|c|}
			\hline
			{Cell radius $d_{\rm max}$} & $ 1000$ m  \\ \hline
			{Reference distance $d_0$} & 100 m\\ \hline
			{Carrier frequency $f_c$}& 3 GHz\\ \hline
			{Channel bandwidth $B_w$}& 20 MHz\\ \hline
			{Pathloss exponent $\gamma$} & 3.71  \\ \hline
			%{Maximal transmitting power} & 316 mW (25 dBm)\\ \hline
			{Reference temperature / Noise figure} & 290\,K / 6\,dB\\ \hline
			{Standard deviation of shadow fading} $\sigma_{\psi}$ & 3.16\\ \hline
		\end{tabular}
	\end{center}
	\label{tbl:simpar} % is used to refer this table in the text
\end{table}

\begin{figure}
	\centering
	\flushleft
	\resizebox{16cm}{!}{\includegraphics{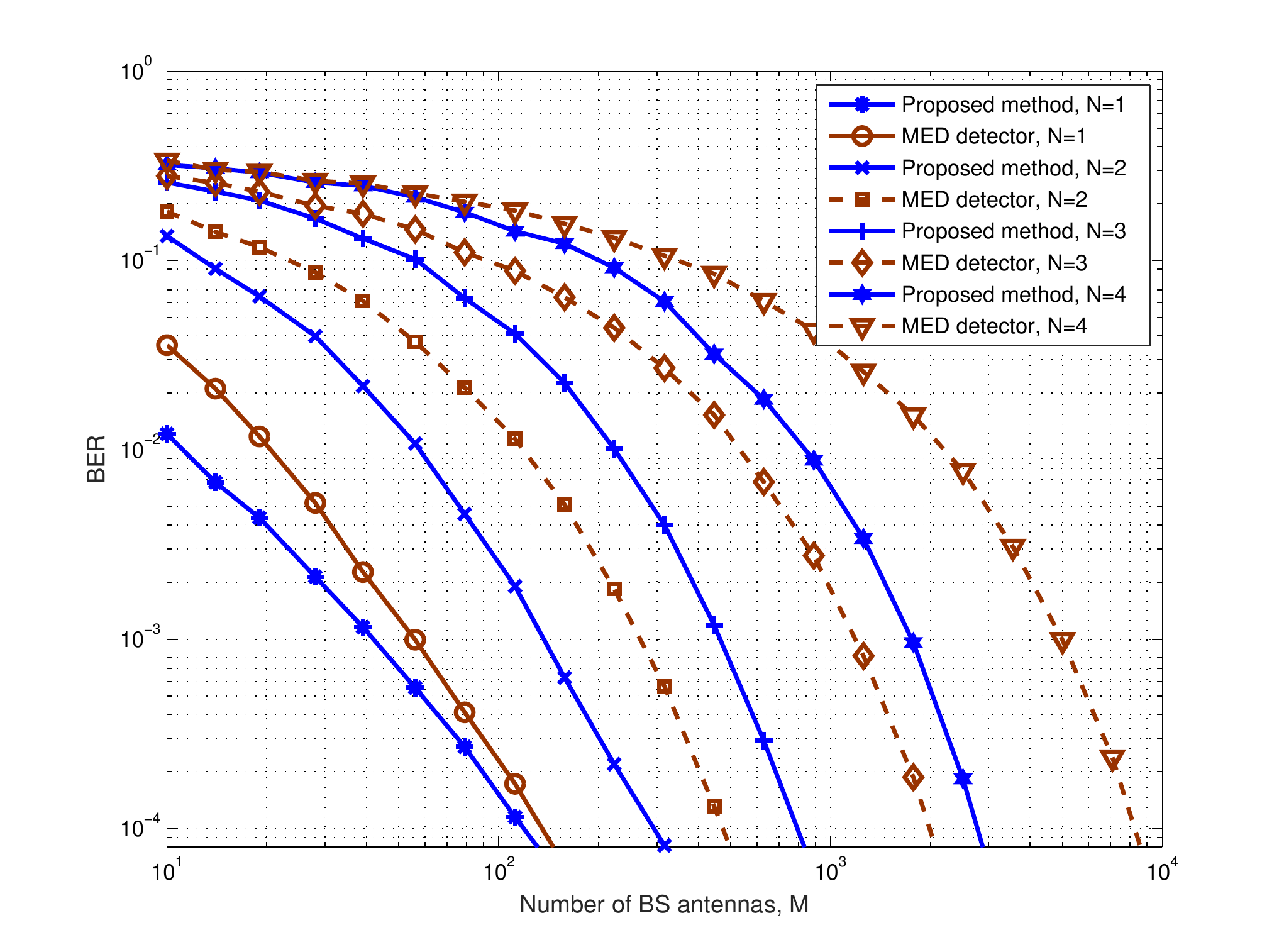}}
	\centering
	\caption{Comparison of the proposed scheme with MED detector on the average BER of all users versus $M$, 4-QAM are used by all the users with average power constraint.}
	\label{fig:avrbervsdist}
\end{figure}

\begin{figure}
	\centering
	\flushleft
	\resizebox{16cm}{!}{\includegraphics{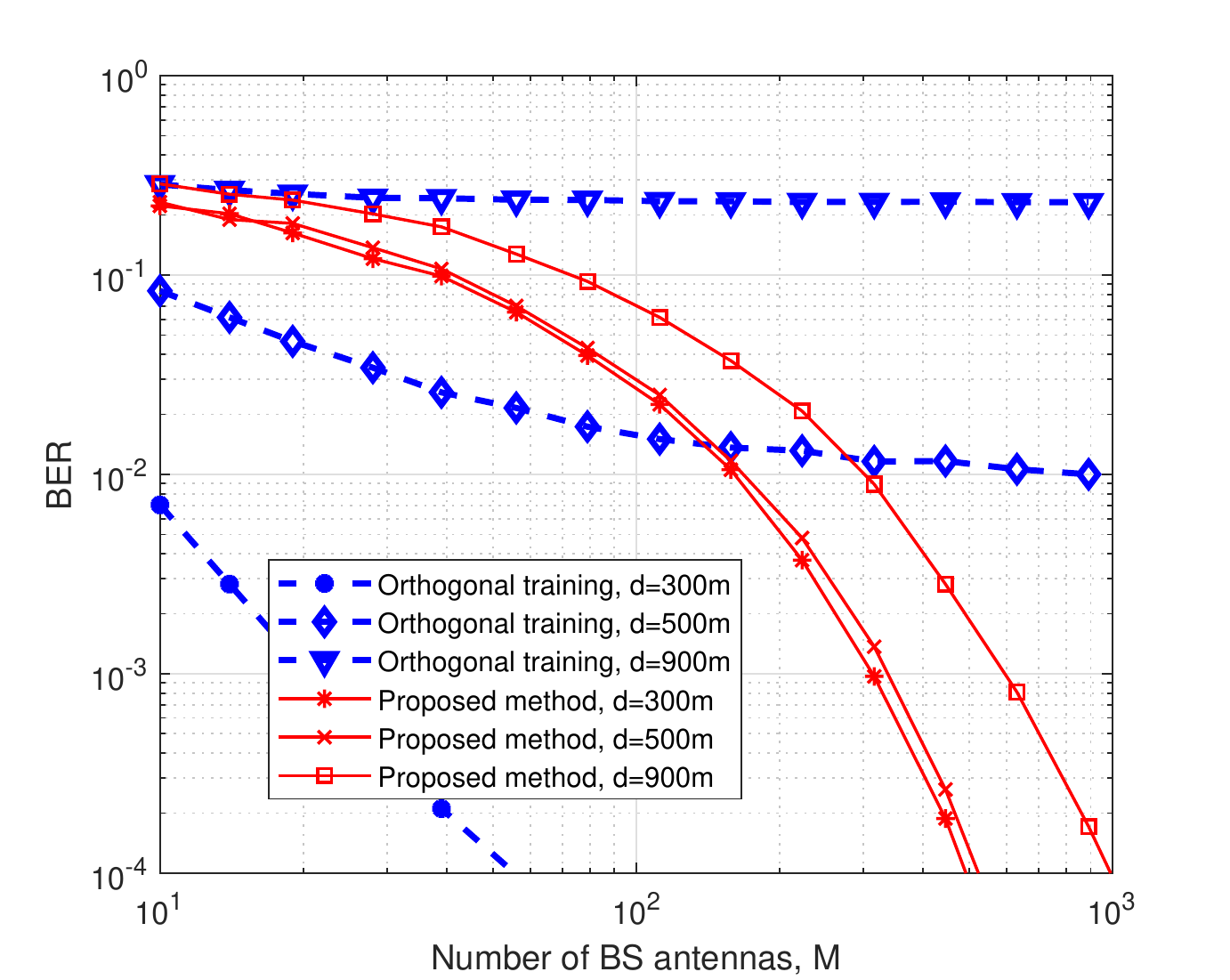}}
	\centering
	\caption{The comparison between the proposed and the orthogonal training method with $N=3$ users and 4 time slot.}
	\label{fig:trainingbsproposed}
\end{figure}

\begin{figure}[ht]
	\centering
	\flushleft
	\resizebox{16cm}{!}{\includegraphics{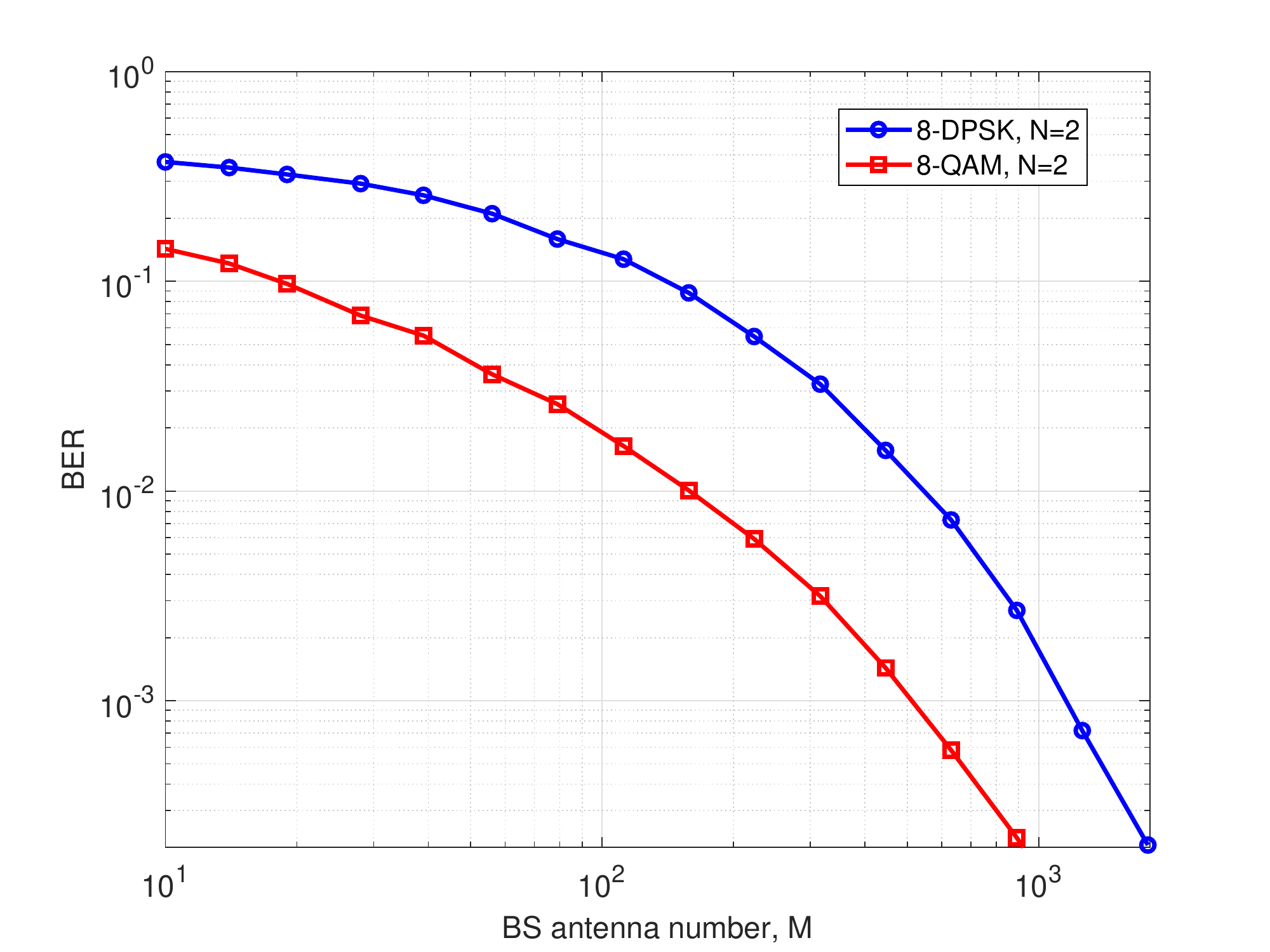}}
	\centering
	\caption{The comparison between the proposed and the noncoherent receiver with 8-QAM and 8-DPSK, respectively.}
	\label{fig:MLvshanzo}
\end{figure}

We first examine the error performance of the proposed design under the instantaneous average power constraint for different number of users, as illustrated in Fig.\,\ref{fig:avrbervsdist}. It is assumed that the average power upper bound is ${P}_k=316$\,mW (25\,dBm), $\forall k$. All the $K$ users are assumed to be uniformly distributed within the cell of radius $d$. It can be observed that, as the number of users increases, the error performance deteriorates quickly which is caused by the mutual interference among users. Then, more BS antennas are needed to achieve the same average BER.  We also compare our  design with the {max-min Euclidean distance (MED)}-based method proposed in~\cite{Goldsmith16tit,Zhang2018JSAC}. Since we use two time slots, while the MED methods only need one time slot, we assume that 2-PAM constellations are adopted by all users for the MED based design. We can see from the figure that the proposed approach significantly outperforms the MED-based method in terms of BER in all simulated cases.

We next compare the error performance of the proposed framework with the conventional zero-forcing (ZF) receiver using orthogonal training sequence. The results are shown in Fig.~\ref{fig:trainingbsproposed}. In this simulation, we consider a system setup with $N=3$ users. For the orthogonal training-based method, at least 4 time slots (3 time slots for training and 1 time slot for data transmission) are needed and we assume that the channel coefficients are quasi-static within these consecutive time slots. As 4-QAM is adopted by each user for the proposed scheme, 64-QAM are correspondingly adopted for the training-based approach in order to make a fair comparison. For the channel training algorithm, we consider that a widely-used least-square (LS) channel estimator is employed~\cite{Gershman06}. It can be observed from Fig.\,\ref{fig:trainingbsproposed} that, when the antenna number $M$ is small and the channel gain is large (i.e., the distance $d$ is small), the training-based method outperforms the proposed design in term of BER. However, when the antenna number is relatively large, the proposed design has a better error performance, especially at the cell edge. The rationale is that \emph{without a reliable CSI, especially at low signal-to-noise ratio (SNR) regimes, coherent detection suffers from inferior decoding performance.}

It is finally worth mentioning that a related noncoherent multiuser massive MIMO system was designed in~\cite{Hanzo15}  for differential phase shift keying (DPSK) constellations. The transmitted information of all the users is modulated on the phase offset between successive symbols. In fact, the DBPSK and the DQPSK constellations with optimal scale between every sub-constellation are the special case of our QAMD. However, for larger constellations such as 8-DPSK, our design has greater normalized minimal Euclidean distance. The resulting sum-constellation of two 8-DQPSK is not a regular constellation anymore, just as studied in~\cite{Harshan11}. Also, in\cite{Hanzo15}, the actual transmitted power of each user is not given explicitly, and hence, the optimal power allocation under both the average and the peak power constraint case is hard to evaluate. To make a comparison, especially when the constellation size is large, we compare the 8-DPSK constellation suggested in~\cite{Hanzo15} with the optimal scale 1.765 between the two sub-constellations with the rectangular 8-QAM constellation in our case. The error performance of~\cite{Hanzo15} and our proposed design with two users, using 8-DPSK and 8-QAM respectively, is studied in Fig.\,\ref{fig:MLvshanzo}. It can be observed that our scheme with 8-QAM sub-constellation has a better error performance than~\cite{Hanzo15} with 8-DPSK constellation, since the normalized minimal distance for our constellation is larger. Also, it should be pointed out that the resulting sum-constellation in~\cite{Hanzo15} is not a regular constellation and it must be either computed or stored in advance. The detection of the sum-constellation typically requires a exhaustive search over the whole constellation. In addition, the optimal power scale for general DPSK needs to be optimized by numerical methods. In contract, our design leads to a regular QAM sum-constellations. Furthermore, the optimal transmit powers of all users and the sub-constellation assignment among them have been provided in closed-form.

\section{Conclusions}
{In this paper, a non-orthogonal and noncoherent massive MIMO (nn-mMIMO) framework towards enabling scalable URLLC applications has been developed based on a new uniquely-factorable multiuser space-time modulation (UF-MUSTM) scheme. For the MUSTM code design, a simple yet systematic construction method based on the concept of QAM division has been devised. Assuming that the large scale fading coefficients are known at the base station, the detailed transmission scheme and the corresponding noncoherent detector have been carefully designed. We further optimized the proposed design framework by jointly optimizing the constellations of multiple users. Specifically, we implemented a max-min Kullback-Leibler (KL) {divergence}-based design criterion, based on which we jointly optimize the transmitted powers of all users and the sub-constellation assignment among them. Simulations demonstrated that the optimized nn-mMIMO framework has better reliability performance compared to the state-of-the-art benchmarking schemes.}

\section*{Appendix}
\begin{appendices}
\subsection{Proof of Proposition~\ref{proposition:UFCM}}\label{append:prop1}
We first show the sufficiency of Proposition~\ref{proposition:UFCM} for the considered massive MIMO system with unlimited number of antennas. By Assumption~\ref{assumption1} on channel statistics and the central limit theory, we have
$\lim_{M\to \infty} \frac{\mathbf{G}^H \mathbf{G} }{M}=\mathbf{I}_K$ and $
\lim_{M\to \infty} \frac{\mathbf{\Xi}^H \mathbf{\Xi}}{M} =\sigma^2\mathbf{I}_T$. Now, the receiver can employ a simple correlation-based detector by calculating $\mathbf{R}_M= \frac{\mathbf{Y}^H \mathbf{Y}}{M}$. When the antenna array size goes to infinity, we have $\lim_{M\to \infty} \mathbf{R}_M -\sigma^2 \mathbf{I}_T=\mathbf{R}$, where ${\mathbf R}$ is a $T\times T$ Hermitian positive semidefinite matrix such that
$\mathbf{R} = \mathbf{X}_T^H \mathbf{D}\mathbf{X}_T$. Now, $\mathbf{X}_T$ can be uniquely determined by an exhaustive search since for any $\mathbf{X}_T, \widetilde{\mathbf{X}}_T\in {\mathcal M}_{K\times T}$ with $\mathbf{X}_T^H \mathbf{D}\mathbf{X}_T= \widetilde{\mathbf{X}}_T^H \mathbf{D}\widetilde{\mathbf{X}}_T$, we have ${\mathbf{X}}_T=\widetilde{\mathbf{X}}_T$.

Next, we show the necessity of Proposition~\ref{proposition:UFCM}. Suppose that there exist $\mathbf{X}_T, \widetilde{\mathbf{X}}_T\in {\mathcal M}_{K\times T}$ such that ${\mathbf{X}}_T \neq \widetilde{\mathbf{X}}_T$ for $ \mathbf{X}_T^H \mathbf{D}\mathbf{X}_T= \widetilde{\mathbf{X}}_T^H \mathbf{D}\widetilde{\mathbf{X}}_T$. As a consequence, ${\mathbf{X}}_T$ and $\widetilde{\mathbf{X}}_T$ will have exactly the same likelihood function as shown in Eq.\,\eqref{eqn:MLdetector}, and hence they are indistinguishable by the ML detector where reliable recovery of the transmitted signals can not be guaranteed. This finish the proof of Proposition~\ref{proposition:UFCM}.\hfill$\Box$

\subsection{Proof of Proposition~\ref{prop:udcg}}\label{append:prop2}
Let ${\mathbf X}_2=\mathbf{D}^{-1/2}{\mathbf S}_2$ and $\widetilde{\mathbf X}_2=\mathbf{D}^{-1/2}\widetilde{\mathbf S}_2$. Then, if ${\mathbf X}_2^H{\mathbf D}{\mathbf X}_2=\widetilde{\mathbf X}_2^H{\mathbf D}\widetilde{\mathbf X}_2$, and note that $\mathbf{\Pi}^H\mathbf{\Pi}=\mathbf{I}_M$, we have ${\mathbf S}_2^H{\mathbf S}_2=\widetilde{\mathbf S}_2^H\widetilde{\mathbf S}_2$. As a consequence, we have
$\sum_{k=1}^N s_k =\sum_{k=1}^K \tilde{s}_k$, where $s_k, \tilde{s}_k\in  \mathcal{X}_k$. Since $\{{\mathcal X}_k\}_{k=1}^K$ form a UDCG, by Lemma~\ref{lemma:UDCG}, we can attain $s_k =\tilde{s}_k$, or equivalently, $\mathbf{S}_2=\tilde{\mathbf{S}}_2$ and now we have ${\mathbf X}_2 =\widetilde{\mathbf X}_2$.

This completes the proof of Proposition~\ref{prop:udcg}. \hfill $\Box$

\subsection{Proof of Lemma~\ref{lemma:orderedseq}} \label{appendix:orderedseq}
%\begin{lemma}\label{lemma:orderedseq}
%	Suppose that there are two ordered real positive sequences $\{a_k| a_k \le a_\ell {~\rm if~} k<\ell, k,\ell=1,2, \ldots, K\}$ and $\{b_k|b_k \le b_\ell {~\rm if~} k<\ell, k,\ell=1,2, \ldots, K\}$.
%	Let $\mathcal{U}=\{ (\pi(1), \pi(2), \ldots, \pi(K)) \}$ be the set containing all possible permutations on $(1,2,\ldots, K)$, in which $(\pi(1), \pi(2), \ldots, \pi(K))$ is a permutation on $(1,2, \ldots, K)$ and $\pi(k)$ is the $k$-th entry of $(\pi(1), \pi(2), \ldots, \pi(K))$.
%	Then, the following problem
%	\begin{align}
%	\max_{ (\pi(1), \pi(2), \ldots, \pi(K)) \in \mathcal{U}} \min_{k=1,2,\ldots,K} \Big\{\frac{a_k}{b_{\pi(k)}}\Big\}
%	\end{align}
%	has a solution $(\pi^*(1), \pi^*(2), \ldots, \pi^*(K))=(1,2, \ldots, K)$.\hfill\QED
%\end{lemma}

%\emph{Proof:}
Let $m=\arg \min_{k=1,2, \ldots, N}\big\{ \frac{a_k}{b_{\pi^*(k)}}\big\}= \arg \min_{k} \big\{\frac{a_k}{b_k}\big\}$. In other words, $m$ is the index such that $q_m=\frac{a_m}{b_m} =\min_{k=1,2,\ldots, N} \big\{\frac{a_k}{b_k}\big\}$. Now, we want to show that $q_m =\max_{(\pi(1), \pi(2), \ldots, \pi(N)) \in \mathcal{U}} \min_{k=1,2,\cdots,N} \big\{\frac{a_k}{b_{\pi(k)}}\big\}$.
To that end, we divide $\mathcal{U}$ into two mutually exclusive subsets, i.e., $\mathcal{P}=\{(\pi(1), \pi(2), \ldots, \pi(N)) |\pi(m) \neq m \}$ and $\mathcal{U} \setminus \mathcal{P} =\{(\pi(1), \pi(2), \ldots, \pi(N))|\pi(m) = m \}$. Consider the following cases:
\begin{itemize}
	\item $ (\pi'(1), \pi'(2), \ldots, \pi'(N)) \in \mathcal{P}$. In this case,
	%we have $\min_{k=1,2, \ldots, K} \big\{\frac{a_k}{b_{\pi'(k)}}\big\}\le q_m$.
	%This fact can be proved by following arguments.
	%As $\pi'(m) \neq m$,
	there exists an $\ell \neq m$ such that $\pi'(\ell)=m$ and hence $b_{\pi'(\ell)} =b_m$.
	%It can be further categorized into the following two cases:
	%\begin{itemize}
		If $\ell < m$, then, we have $\frac{a_\ell}{b_{\pi'(\ell)}}= \frac{a_\ell}{b_{m}} \le \frac{a_m}{b_m} =q_m$.
		If $\ell > m$, there exits an $n \le m$ such that $\pi'(n) > m$ by the property of permutation.
		Then, we have $\frac{a_n}{b_{\pi'(n)}}\le \frac{a_m}{b_{\pi'(n)}} \le \frac{a_m}{b_m}=q_m$.
	%\end{itemize}
	Therefore, we conclude $\min_{k=1,2,\ldots, N} \big\{\frac{a_k}{b_{\pi'(k)}}\big\} \le q_m$ for any $ (\pi'(1), \pi'(2), \ldots, \pi'(N)) \in \mathcal{P}$. Or equivalently, $\max_{ (\pi(1), \pi(2), \ldots, \pi(N)) \in \mathcal{P}} \min_{k=1,2,\cdots, N} \big\{\frac{a_k}{b_{\pi(k)}}\big\} \le q_m$.
	%Specifically, it can be observed that if $0<a_1<a_2<\ldots<a_K$ and $0<b_1<b_2<\ldots<b_K$ , we have $\max_{ (\pi(1), \pi(2), \ldots, \pi(K)) \in \mathcal{P}} \min_{k=1,2,\ldots,K} \big\{\frac{a_k}{b_{\pi(k)}}\big\} < q_m$.
	\item $(\pi'(1), \pi'(2), \ldots, \pi'(N)) \in \mathcal{U} \setminus \mathcal{P}$.  In this case, $\pi'(m)=m$ and hence, we have $\min_{k=1,2,\cdots, N} \big\{\frac{a_k}{b_{\pi'(k)}}\big\} \le \frac{a_m}{b_{\pi'(m)}}=\frac{a_m}{b_m}=q_m$. Therefore, $\max_{ (\pi(1), \pi(2), \cdots, \pi(N)) \in \mathcal{U} \setminus \mathcal{P}} \min_{k=1,2,\cdots,N} \big\{\frac{a_k}{b_{\pi(k)}}\big\} \le q_m$.
\end{itemize}
In conclusion, we have $\max_{\pi \in \mathcal{U}} \min_{k=1,2,\cdots, N} \big\{\frac{a_k}{b_{\pi(k)}}\big\} \le q_m$.
In the following, we aim to prove that the equality is achievable for certain $(\pi(1), \pi(2), \ldots, \pi(K))$.
By setting $(\pi(1), \pi(2), \cdots, \pi(N)) =(\pi^*(1), \pi^*(2), \ldots, \pi^*(N))$ and then, from the construction process above, we can find that for the given sequences $a_1\le a_2\le \cdots \le a_N$ and $b_1\le b_2\le \cdots \le b_N$, $\min_{k=1,2, \cdots, N} \big\{\frac{a_k}{b_{\pi^*(k)}}\big\} =\frac{a_m}{b_m}= q_m$.
Hence, the equality is achievable for $(\pi^*(1), \pi^*(2), \cdots, \pi^*(N))$.
%It is worth noting that for $a_1< a_2< \cdots < a_N$ and $b_1<b_2<\cdots <b_N$, $\pi(m)=m$ is a necessary condition to attain the lower bound on $\max_{ (\pi(1), \pi(2), \ldots, \pi(N)) \in \mathcal{U}} \min_{k=1,2,\ldots, N} \big\{\frac{a_k}{b_{\pi(k)}}\big\}$.
This completes the proof.~\hfill$\Box$

\end{appendices}
\small
\bibliographystyle{ieeetr}
\bibliography{reference}

\normalsize

% ----- Document End -----------------------------------------------------

\end{document}